\documentclass[preprint]{elsarticle}
\pdfoutput=1
\usepackage{amsmath}
\usepackage{amssymb}
\usepackage[final]{epsfig}
\usepackage{color}
\usepackage{graphicx}
\usepackage{lineno}
\newcommand{\beq}{\begin{equation}}
\newcommand{\eeq}{\end{equation}}
\newcommand{\bdi}{\begin{displaymath}}
\newcommand{\edi}{\end{displaymath}}
\newcommand{\no}{\nonumber}
\newcommand{\bea}{\begin{eqnarray}}
\newcommand{\eea}{\end{eqnarray}}

\newcommand{\ov}{\overline}

\begin{document}

\begin{frontmatter}

\title{Time resolution of silicon pixel sensors}

\author[]{W. Riegler} 
\author{G. Aglieri Rinella}
%\corref{cor1}

\address{CERN EP, CH-1211 Geneve 23}
%\ead{werner.riegler@cern.ch}

\begin{abstract}

We derive expressions for the time resolution of silicon detectors, using the Landau theory and a PAI model for describing the charge deposit of high energy particles. First we use the centroid time of the induced signal and derive analytic expressions for the three components contributing to the time resolution, namely charge deposit fluctuations, noise and fluctuations of the signal shape due to weighting field variations. Then we derive expressions for the time resolution using leading edge discrimination of the signal for various electronics shaping times. Time resolution of silicon detectors with internal gain is discussed as well. 

\end{abstract}


\end{frontmatter} 

\newpage
\tableofcontents
\clearpage

\section{Introduction}

Silicon pixel sensors providing precise timing are currently being
developed in view of future "4D" tracking applications. 
The NA62 Gigatracker, using sensors of 200\,$\mu$m thickness and
$300\,\mu$m$\times 300\,\mu$m pixel size has achieved time resolutions
of $\le 150$\,ps at rates of up to 1.5\,MHz/cm$^2$
\cite{AglieriRinella2017147}\cite{Rinella20121608}\cite{Kluge2013511}\cite{Fiorini2013270}. 
A time resolution of 100\,ps has been reported with a sensor of 100\,$\mu$m thickness and $800\,\mu$m$\times
800\,\mu$m pixel size \cite{100ps_timing}. For multiple particles passing silicon sensors of thickeness between 133 and 285\,$\mu$m, a time resolution of better than 20\,ps has been reported \cite{Akchurin2017}.
With the introduction of internal amplification inside silicon detectors of 50\,$\mu$m
thickness, the so called Low Gain Avalanche Diode (LGAD) 
\cite{Pellegrini201412}\cite{Cartiglia2014}\cite{Sadrozinski2013226}\cite{Sadrozinski20147}\cite{Cartiglia2015141},
time resolutions of 25\,ps have been achieved for single MIPs \cite{Cartiglia201783}. \\ \\ 
The Weightfield2 program \cite{Cenna2015149} allows the detailed
simulation of the induced signals in silicon sensors with strip
geometry.  A long term goal of these developments are pixel sensors of
10\,$\mu$m position resolution and $10$\,ps time resolution
\cite{Cartiglia201747}\cite{Sola-1748-0221-12-02-C02072}. 
Developments of silicon sensors for increased timing performance 
based on 3D sensors are also described in literature \cite{Parker-5733383}.
Studies of front-end electronics for silicon detectors with emphasis on 
timing aspects can be found in \cite{Spieler-4336333}  and \cite{Rivetti2014202}. 
Charged particle imaging is widely employed in many areas of science beyond 
high energy physics, for example as part of material analysis techniques. 
Therefore there is a broad interest in the developments of spatially resolved 
and time accurate particle detectors \cite{Jungmann20125077}\cite{Vallance2014383}.\\ \\
In this report we derive analytic expressions for the time resolution of
silicon sensors using the Landau theory and a version of the PAI model to describe 
the charge deposit of high energy particles in the sensor.  We first investigate the
time resolution for the case where we take the 'centroid
time' of the signal as a measure of time. It refers to the case where
the amplifier peaking time is larger than the drift time of the
electrons and holes in the silicon sensor and allows us to discuss the
achievable time resolution using moderate electronics bandwidth
together with optimum filter methods to extract the time information
from the known signal shape. We then derive formulas quantifying the
effect of signal fluctuations due to the finite pixel size and related
variations of the weighting field.  We also derive expressions for
the time resolution using leading edge discrimination of the signals
with different electronics shaping times. In the last part of the report we discuss the time resolution of silicon sensors with internal amplification which will be applied in the ATLAS and CMS experiment upgrades for pileup rejection \cite{Cartiglia2014}.

\section{Energy deposit}

\begin{figure}
 \begin{center}
   a)
   \includegraphics[width=6cm]{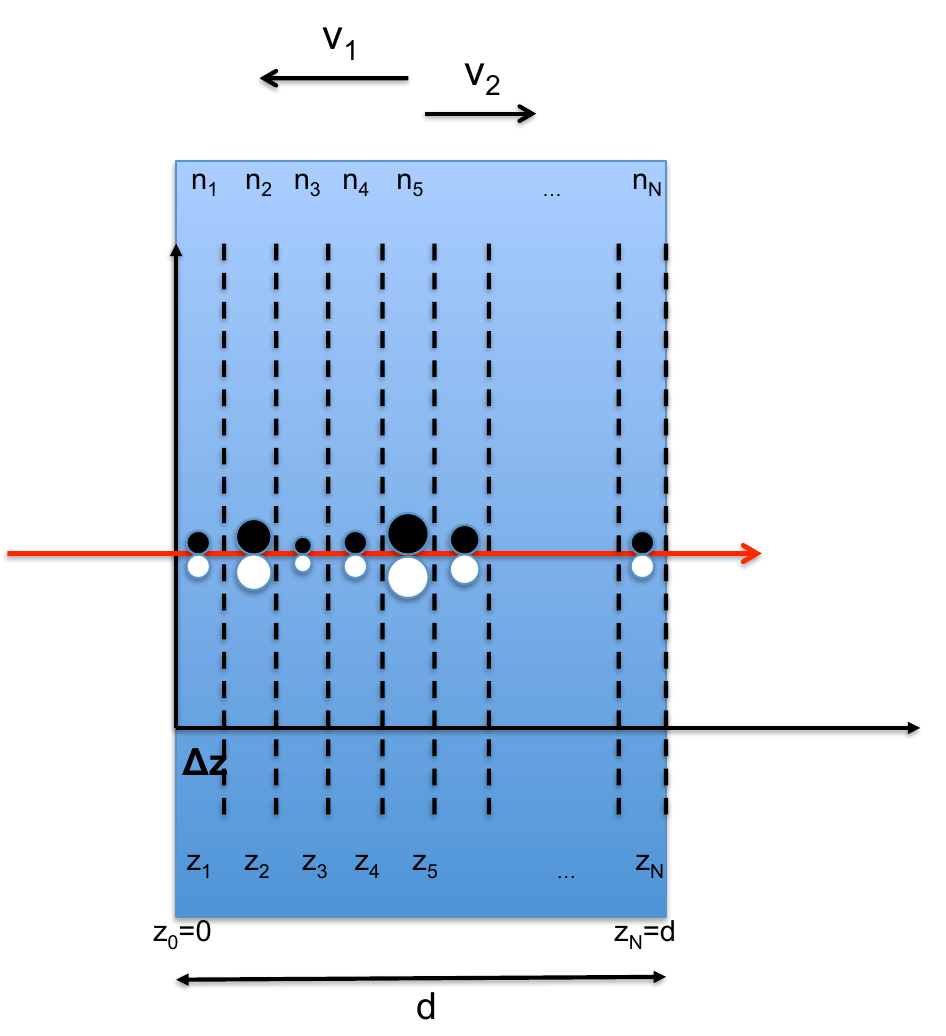}
   b)
   \raisebox{+0.2\height}{ \includegraphics[width=7cm]{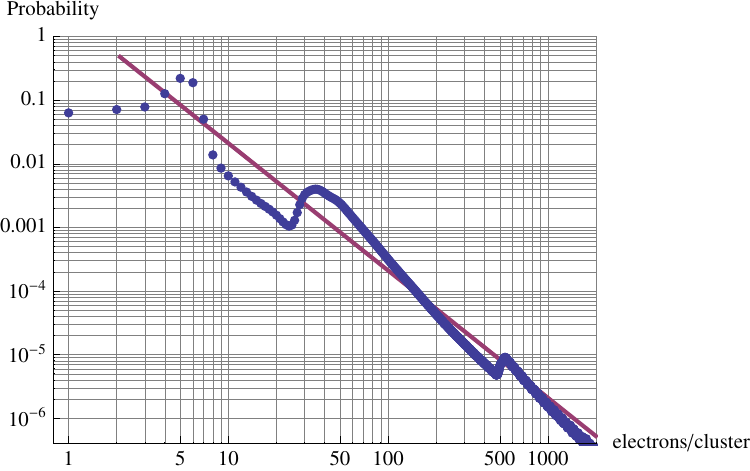}}
  \caption{a) The silicon sensor is divided into slices of thickness $\Delta z$. The electrons and holes produced in one slice are assumed to move to the boundary of the sensor at constant velocity, which is correct in the limit of negligible depletion voltage. b) Probability to find $n$ electrons per primary interaction. The straight line refers to the $1/n^2$ distribution that is the basis for the Landau distribution, the points corresponds to a PAI model \cite{heinrich}\cite{PAI}. }
  \label{sensor_slices}
  \end{center}
\end{figure}
\begin{figure}
 \begin{center}
   a)
   \includegraphics[width=6cm]{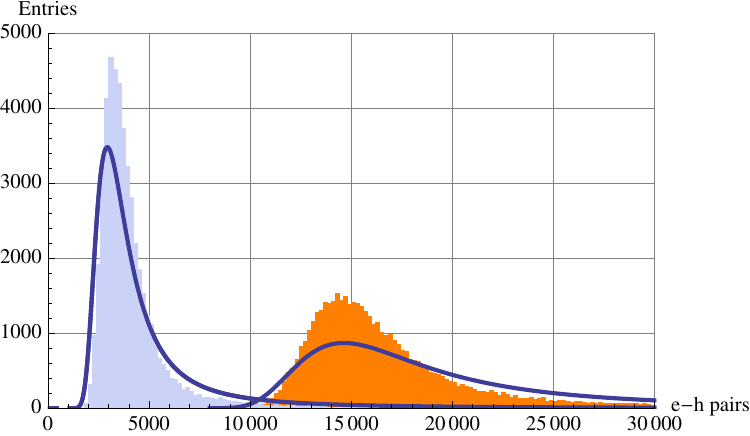}
   b)
   \includegraphics[width=7cm]{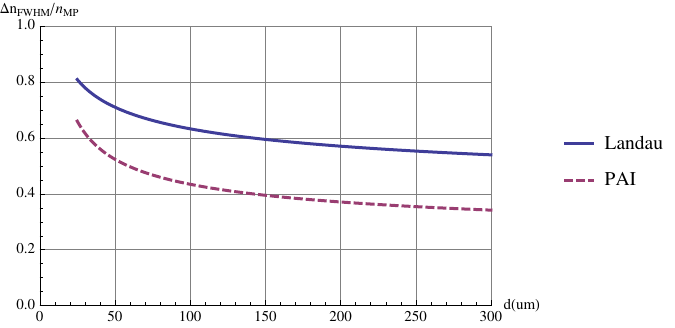}
  \caption{a) Distribution of the number of e-h pairs in 50\,$\mu$m (blue) and 200\,$\mu$m (orange) of silicon. The histograms show the PAI model, the solid lines show the Landau theory. b) Ratio of full width half maximum and most probable values for the Landau and PAI model for different values of silicon thickness.  The Landau theory overestimates the fluctuations by 25-35\%.}
  \label{landau}
  \end{center}
\end{figure}

A high energy particle passing a silicon sensor will experience a number of primary interactions with the material, with $\lambda$ being the average distance between these primary interactions. For relativistic particles we have $\lambda \approx 0.212\,\mu$m in silicon \cite{heinrich}. The electrons created in these primary interactions will typically lose their energy over very small distances and create a localised cluster of electron-hole pairs. We call the probability $p_{clu}(n)$ for creating $n$ e-h pairs in a primary interaction the 'cluster-size distribution'. Throughout this report we treat $n$ as a continuous variable. 
We now divide the silicon sensor of thickness $d$ into $N$ slices of thickness $\Delta z = d/N$ as shown in Fig. \ref{sensor_slices}a. In case $\Delta z \ll \lambda$, the probability for having zero interactions in $\Delta z$ is $1-\Delta z/\lambda$, the probability to have one interaction in $\Delta z$ is $\Delta z/\lambda$ and the probability to have more than one interaction is negligible, so the probability density for finding $n$ electrons in $\Delta z$ is 
\beq \label{pn_approximation}
    p(n, \Delta z)dn = \left(1-\frac{\Delta z}{\lambda} \right)  \delta(n) dn + 
    \frac{\Delta z}{\lambda} p_{clu}(n)dn
\eeq 
The probability $p(n,d)$ to have $n$ electrons in the entire sensor of thickness $d$ is then given by the $N$ times self convolution of this expression. Since convolution becomes multiplication if we perform the Laplace transform, $N$ times self convoluting the above expression results in raising it's Laplace transform to the power $N$. So using the Laplace transform $P_{clu}(s)={\cal L}[p_{clu}(n)]$ we have 
\beq
   P(s, d) = {\cal L}[p(n, d)] = {\cal L} [p(n,\Delta z)]^N
   =\left(
   1+\frac{d}{\lambda N}(P_{clu}(s)-1)
   \right)^N
\eeq
By taking the limit of $N \rightarrow \infty$ we have 
\beq \label{general_pn_distribution}
   p(n, d) = {\cal L}^{-1}
   \left[
   e^{d/\lambda(P_{clu}(s)-1)} 
   \right]
\eeq
This expression is completely general and correct for any cluster size distribution. Assuming as an (unphysical) example that each cluster contains exactly $n_0$ electrons we have
\beq
    p_{clu}(n)=\delta(n-n_0) \qquad 
    P_{clu}(s)=e^{-sn_0} \qquad 
    P(n,s) = e^{d/\lambda(e^{-n_0s}-1)} 
\eeq
The inverse Laplace transform of the last expression is
\beq \label{poisson_expample}
   p(n,d) = \sum_{k=0}^\infty \frac{\left( \frac{d}{\lambda} \right) ^k}{k! } 
   e^{-\frac{d}{\lambda}}
   \delta(n-k\,n_0)
   \qquad
   \mu = n_0d/\lambda
   \qquad
    \frac{\Delta }{\mu} = \frac{1}{\sqrt{d/\lambda}}
\eeq
where $\mu$ is the average number of e-h pairs and and $\Delta$ is the standard deviation. This is the expected Poisson distribution showing the $1/\sqrt{N}$ dependence for the relative fluctuations with $N=d/\lambda$ being the average number of clusters. \\
The correct cluster size distribution $p_{clu}(n)$ is typically calculated using some form of the PAI model \cite{PAI} and an example is shown in Fig. \ref{sensor_slices}b \cite{heinrich}. For this report we also use the Landau theory as a minimal model that respects basic physics and that allows approximate analytic expressions. Landau's approach assumes a $1/E^2$ distribution of the energy transfer for a collision in accordance with Rutherford scattering on free electrons and a lower cutoff energy $\epsilon$ chosen such that the average energy loss reproduces the Bethe-Bloch theory. The resulting cluster size distribution for a MIP in silicon therefore becomes a $1/n^2$ distribution with a cutoff at $n=n_0 \approx 2.2$\,electrons, which can be written as
\beq \label{Landau_approximation}
    p_{clu}(n) \approx  \frac{n_0}{n^2}\,\Theta(n-n_0)
    \qquad
    P_{clu}(s) \approx 1+n_0s(C_{\gamma}-1+\ln n_0+ \ln s)
\eeq 
with $\Theta(x)$ being the Heaviside step function. Evaluating Eq. \ref{general_pn_distribution} results in 
\beq \label{landau_probability}
    p(n,d)dn = \frac{\lambda}{n_0\,d}\, 
    L\left( 
    \frac{\lambda}{n_0\,d}\,n+C_\gamma-1-\ln \frac{d}{\lambda}
     \right) \, dn
\eeq
where $C_\gamma = 0.5772...$ is the Euler-Mascheroni constant and $L(x)$ is the Landau distribution discussed in \ref{appendix_landau}. The most probable number of e-h pairs $n_{MP}$ and the full width of half maximum $n_{FWHM}$ of $p(n, d)$ are
\beq \label{mp_fwhm}
       n_{MP} \approx \frac{n_0\,d}{\lambda}
     \left(
     0.2+\ln \frac{d}{\lambda}
     \right)
     \qquad
\frac{\Delta n_{FWHM}}{n_{MP} }\approx \frac{4.02}{0.2+\ln d/\lambda}
\eeq
It should be noted that the most probable number of electrons $n_{MP}$ is proportional to the cutoff $n_0$ while the ratio of $n_{FWHM}$ and $n_{MP}$ is independent of $n_0$ and depends only on $d/\lambda$. \\
For a value of $\lambda = 0.212\,\mu$m we find an average of $N=d/\lambda =236, 472, 943, 1415$ primary interactions (clusters) for a $50, 100, 200, 300\,\mu$m silicon sensor. Using the cluster size distribution from Eq. \ref{Landau_approximation}, the probability that at least one of the $N$ clusters contains more than $n_1$ electrons is given by 
\beq
    p_{>n_1}=1-\left(1-\frac{n_0}{n_1}\right)^N
\eeq
so there is still a 1\% chance to have a cluster with more than $n_1=73500, 103000, 206000, 309000$ electrons for a single MIP passing a 50, 100, 200, 300\,$\mu$m silicon sensor! When performing Monte Carlo simulations, the cut-off of the cluster size distribution has therefore to be placed beyond these numbers. The primary electrons producing these large clusters are called delta-electrons and do not deposit their charge at point-like clusters but short tracks, which has to be considered when discussing pixels of small size. \\
Fig. \ref{landau}a shows the distribution of e-h pairs in a 50\,$\mu$m and a 200\,$\mu$m sensor for the PAI model together with the curves from the Landau theory. As seen in Fig. \ref{landau}b the Landau theory overestimates the fluctuations by 25-35\%.
The PAI model predicts a most probable number of  $3160, 6710, 14200, 21900$ e-h pairs in $50, 100, 200, 300\,\mu$m of silicon, which is within 10\% of the values from the Landau theory when assuming a cutoff of $n_0=2.2$. We will use both models for evaluation of the time resolution in the following.

\clearpage

%%%%%%%%%%%%%%%%%%%%
%%%%%%%%%%%%%%%%%%%%%
%%%%%%%%%%%%%%%%%%%%%

\section{Centroid time of a signal}

\begin{figure}
 \begin{center}
   a)
   \includegraphics[width=6cm]{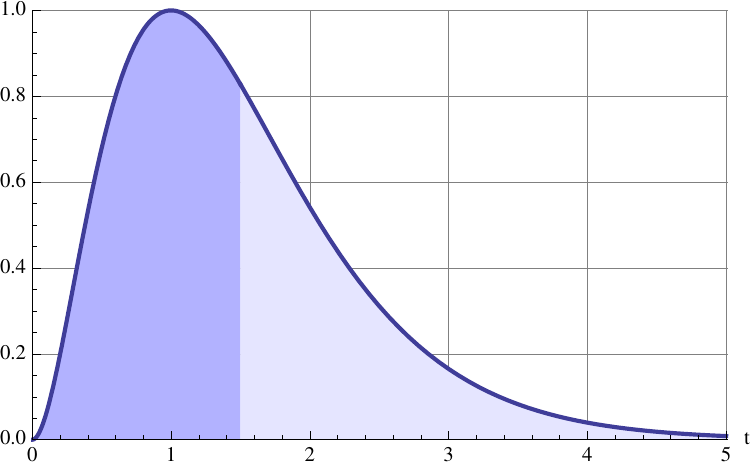}
   b)
   \includegraphics[width=6cm]{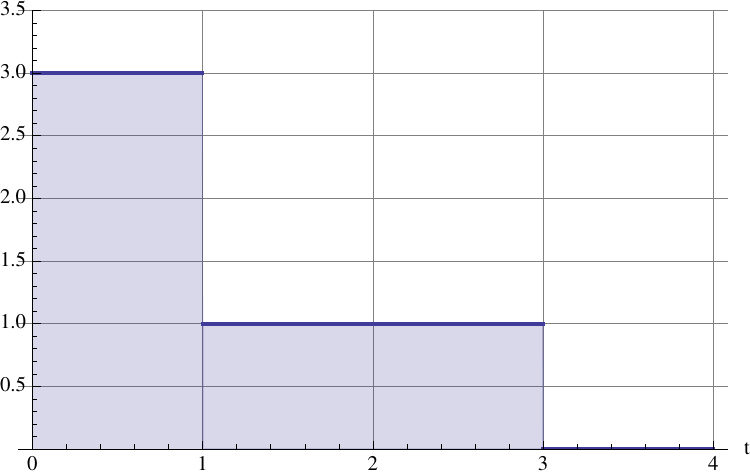}
  \caption{a) The centroid  time $\tau$ of a signal is defined as the time where signal integral (area) for $t<\tau$ and $t>\tau$ is balanced. b) Example of the signal from a single e-h pair in a silicon sensor with negligible depletion voltage.  }
  \label{cog_signal_example}
  \end{center}
\end{figure}

First we assume the measured time to be defined by the centroid time of the induced detector current signal $i(t)$ (Fig. \ref{cog_signal_example}a. Assuming the Laplace Transform of the signal $I(s)={\cal L}[i(t)]$, the centroid  time $\tau_{cur}$ of the signal is defined by 
\beq \label{cogtime}
  \tau_{cur} = \frac{\int_0^\infty t \, i(t) dt}{\int_0^\infty  i(t) dt} = 
  \frac{\int_0^\infty  t \, i(t) dt}{q}= -\frac{I'(0)}{I(0)}
\eeq
where $q= \int_0^\infty  i(t) dt$ is the total signal charge. If we consider the signal $i(t)$ to be processed by an amplifier having a delta response $f(t)$ with Laplace Transform $F(s)$, the amplifier output signal $v(t)$ is given by
\beq \label{convolution}
  v(t)=\int_0^t f(t-t')i(t')dt' \qquad V(s)=F(s)I(s)
\eeq
The centroid  time of the output signal is then 
\beq
  \tau_v = -\lim_{s\rightarrow 0}\frac{V'(s)}{V(s)} = -\frac{F'(0)I(0)+F(0)I'(0)}{F(0)I(0)}=
   -\frac{F'(0)}{F(0)}-\frac{I'(0)}{I(0)} = \tau_{amp}+\tau_{cur}
\eeq
This represents the sum of the centroid  time of the delta response and the one from the current signal, and since the shape of the delta response does not vary in time, the centroid  time variation of the
amplifier output signal is equal to the centroid  time variation of the original input signal and has no dependence on the amplifier characteristics. \\ 
To determine $\tau$ by recording the signal shape and performing the integral of Eq. \ref{cogtime} is not very practical, it is easier to simply process the signal with an amplifier that is 'slow' compared to the signal duration, as shown in the following. In case the duration $T$ of the signal $i(t)$ is short compared to the 'peaking time'  $t_p$ of the amplifier ($i(t)=0$ for $t>T \ll t_p$) we can approximate Eq. \ref{convolution} for $t>T$ according to 
\bea \label{cog_delta}
  v(t) & = &\int_0^T f(t-t')i(t')dt'   \approx  \int_0^T \left[ f(t)-f'(t)t' \right]\,i(t')dt' \no \\ 
        &=& q  \left[ f(t)-  f'(t)\frac{ \int_0^T t' i(t')dt'}{  q  } \right] 
           =  q  \left[ f(t)-  f'(t)\tau_{cur}  \right] \no \\
    & \approx & q\, f(t-\tau_{cur}) 
\eea
The amplifier output is simply equal to the amplifier delta response shifted by the centroid  time of the current signal and scaled  by the total charge of the signal. Since the shape of the amplifier output signal is always equal to the amplifier delta response, we can determine the signal centroid  time either by the threshold crossing time at a given fraction of the signal or by sampling the signal and fitting the known signal shape to the samples. For later use we remark that for the sum of two current signals $i(t)=i_1(t) + i_2(t)$ with centroid  times $\tau_1$ and $\tau_2$ we have
\beq
  \tau =\frac{\int t i(t)dt}{\int i(t) dt} = \frac{\tau_1 \int i_1(t) dt +\tau_2\int i_2(t) dt }{\int i_1(t) dt+\int i_2(t) dt} =
  \frac{\tau_1q_1 +\tau_2 q_2 }{q_1+ q_2}
\eeq
The centroid  time for the sum of $N$ signals $i(t)=\sum_{k=1}^N i_k(t)$ is therefore given by
\beq
        \tau = \frac{1}{\sum_{k=1}^N q_k}\,\sum_{k=1}^N q_k\,\tau_k
\eeq
where $q_k$ and $\tau_k$ are the charges and centroid times of the individual signals $i_k(t)$.

\section{Silicon sensors without internal gain}

\subsection{Centroid time resolution of a silicon detector signal}

\begin{figure}
 \begin{center}
   a) 
   \includegraphics[width=8cm]{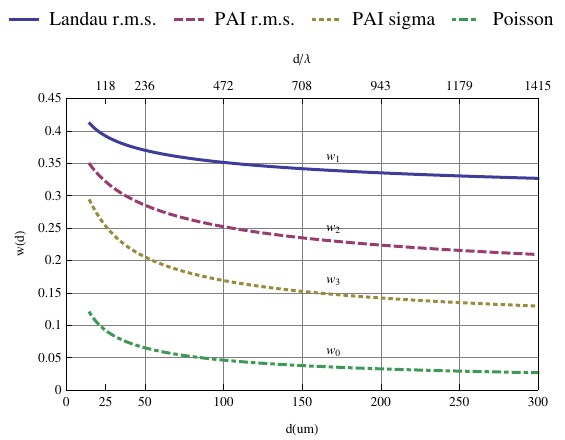}
    b) 
   \includegraphics[width=6cm]{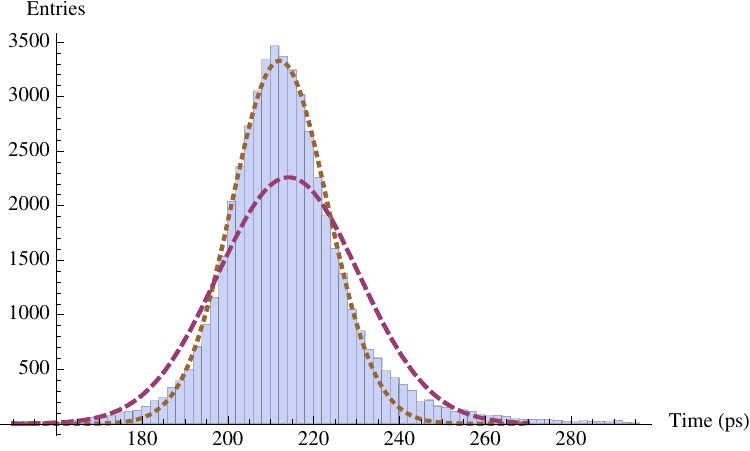}
  \caption{a) The function $c(d)$ for different values of silicon thickness. $w_1$ represents the Landau theory, $w_2$ represents the PAI model and $w_3$ applies for the PAI model if we use a Gaussian fit instead of the r.m.s. as a measure of the time resolution. b) Centroid  time distribution for $d=50\,\mu$m and $V=220\,$V for the PAI model. The dashed curve represents a Gaussian with a $\sigma= \Delta_{\tau}$ ($w_2$) and the dotted curve is a Gaussian fit to the histogram ($w_3$).   }
  \label{cfunction}
  \end{center}
\end{figure}

\begin{figure}
 \begin{center}
 a)
  \includegraphics[width=6cm]{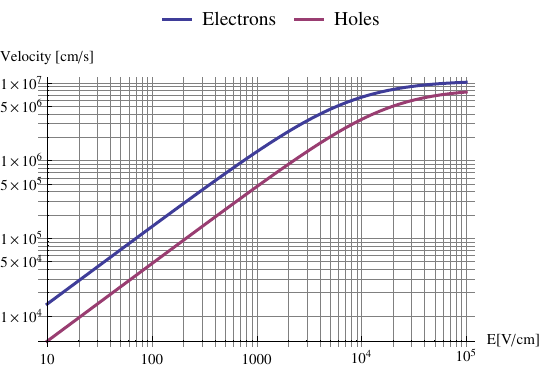}
  b)\includegraphics[width=9cm]{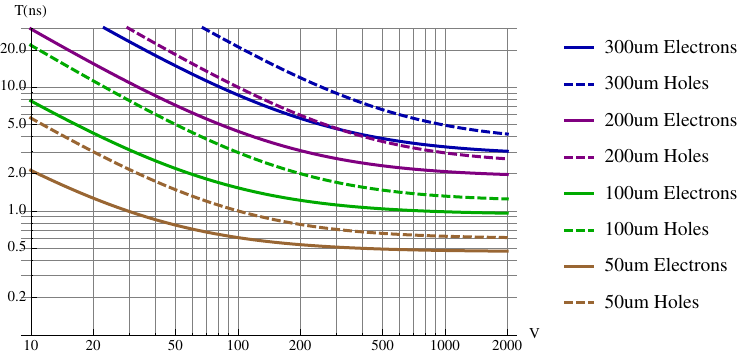}
  \caption{a)Velocity of electrons and holes as a function of electric field.  b) Time for electrons an holes to transit the full thickness of the sensor assuming negligible depletion voltage.}
  \label{velocity}
  \end{center}
\end{figure}

\begin{figure}
 \begin{center}
  \includegraphics[width=14cm]{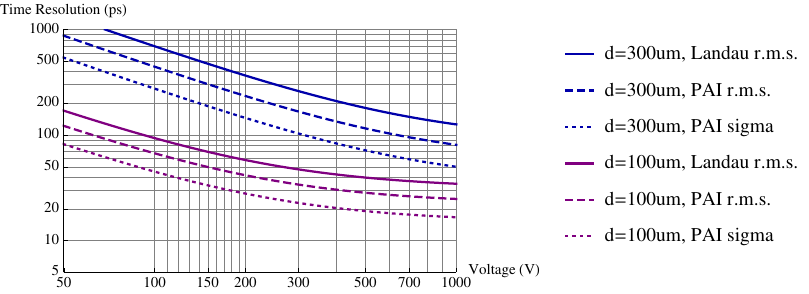}
    \includegraphics[width=14cm]{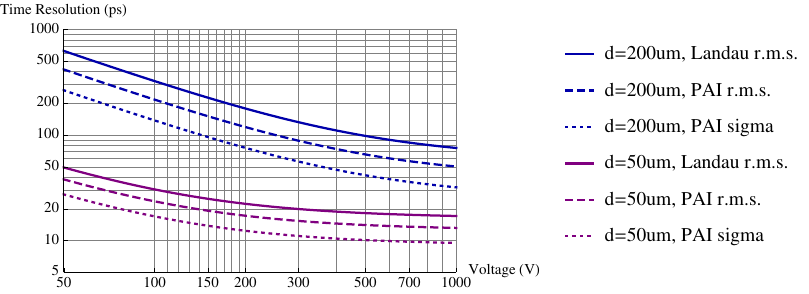}
  \caption{Time resolution from Eq. \ref{final_formula} for different values of silicon sensor thickness as a function of applied voltage $V$ for the Landau model, the PAI model and a Gaussian fit to the PAI model results.  }
  \label{reso}
  \end{center}
\end{figure}

We assume a silicon sensor operated at large over-depletion i.e. at a voltage that is large compared to the depletion voltage and the electric field can therefore be assumed to be constant throughout the sensor. Consequently the velocities of electrons and holes are constant and the signal from a single electron or single hole has a rectangular shape.  
We assume a parallel plate geometry with one plate a $z=0$ and one at $z=d$, where a pair of charges $+q,-q$ is produced at position $z$ and $-q$ moves with velocity $v_1$ to the electrode at $z=0$ while $q$ moves with velocity $v_2$ to the electrode at $z=d$. The weighting field of the electrode at $z=0$ is $E_w=1/d$ and the induced current is therefore 
\beq
      i(t)=-\frac{qv_1}{d} \Theta(z/v_1-t) -\frac{qv_2}{d} \Theta((d-z)/v_2-t) 
\eeq 
with $\Theta(t)$ being the Heaviside step function. An example is shown in Fig. \ref{cog_signal_example}b. We have $\int i(t) dt=-q$ and according to Eq. \ref{cogtime} the centroid  time of this signal is then 
\beq
    \tau = \frac{1}{2d}\left[  \frac{z^2}{v_1}+\frac{(d-z)^2}{v_2}\right]
\eeq
If $n_1, n_2, ..., n_N$ charges are produced at positions $z_1, z_2, ..., z_N$ and are moving to the electrodes with $v_1$ and $v_2$, the resulting  centroid  time of the signal is 
\beq \label{tau_of_n}
   \tau(n_1, n_2, ..., n_N) = 
   \frac{1}{2d\,(\sum_{k=1}^N n_k)}\,
   \sum_{k=1}^N n_k
   \left[
   \frac{z_k^2}{v_1}+\frac{(d-z_k)^2}{v_2}
   \right]
\eeq
We now divide the sensor of thickness $d$ into $N$ slices of $\Delta z = d/N$ as shown in Figure \ref{sensor_slices}. The probability to have $n_k$ e/h pairs in slice $k$ is given by the Landau distribution $p(n_k,\Delta z)$ and if we assume that all these charges are moving from position $z_k$ to the electrodes, we have $z_k=k\,\Delta z$ and we can proceed to calculate the variance $\Delta_{\tau}^2$ of the centroid  time of the signal, i.e. the time resolution, according to 
\beq
    \Delta_{\tau}^2=\ov{\tau^2} - \ov \tau ^2
\eeq
with $\ov \tau$ and $\ov{\tau ^2}$ being the average and the second moment of $\tau$. The evaluation is given in \ref{appendix_statistics} and we find  
\beq \label{final_formula}
   \Delta_{\tau} =
   w(d)\, 
   \sqrt{
   \frac{4}{180}\frac{d^2}{v_1^2}-\frac{7}{180}\frac{d^2}{v_1v_2}+\frac{4}{180}\frac{d^2}{v_2^2}
   }
\eeq
with
\beq
     w(d)^2 = 
              \frac{d}{\lambda}\int_0^\infty 
   \left[ \int_0^\infty \frac{n_1^2\,p_{clu}(n_1)}{(n_1+n)^2}dn_1 \right]
   p(n,d) dn 
\eeq
We first evaluate the expression for the (unphysical) case where we assume each cluster to have exactly $n_e$ electrons i.e. $p_{clu}(n) = \delta(n-n_e)$. The expression inside the square brackets then evaluates to $n_e^2/(n_e+n)^2$. The probability $p(n,d)$ to find $n$ electrons in $d$ is the Poisson distribution from Eq. \ref{poisson_expample} with it's peak at $n=Nn_e$. Since the above expression does not vary significantly within the width of the Poisson distribution, the integral can be approximated by evaluating the expression at $n=N\,n_e$, and we have
\beq
    w(d) \approx \sqrt{
   N\,\frac{n_e^2}{(n_e+N n_e)^2}     }
   \approx \frac{1}{\sqrt {N}} = \frac{1}{\sqrt {d/\lambda}} 
\eeq
This is a very intuitive result related to the typical behaviour of the relative fluctuation of the Poisson distribution.
The evaluation of $w(d)$ for the Landau theory is given in \ref{appendix_wd} with the result that for large values of $d/\lambda$ we have
\beq
      w(d) \approx \frac{1}{\sqrt{\ln d/\lambda} }
\eeq
The value of $w(d)$ is given in Fig. \ref{cfunction}a for the Poisson case $(w_0)$, the Landau theory ($w_1$), the PAI model ($w_2$) and for the case where we do not use the r.m.s. value but a Gaussian fit to the measured times as a measure of the time resolution ($w_3$). As shown in Fig. \ref{cfunction}b the time distribution has very large tails, so the r.m.s. and a Gaussian fit differ significantly. The three curves $w_1, w_2, w_3$ are  parametrized in the range of $15\,\mu$m$<d<300\,\mu$m as 
\beq
   w(d) \approx \frac{1}{\sqrt{a+b\,\ln d/\lambda+c\,(\ln d/\lambda)^2}}
\eeq
with $a_1= 1, b_1=1.155, c_1=0$, $a_2=13.7, b_2=-4.9, c_2=0.85$, $a_3=47.7, b_3=-22.8, c_3=3.37$.
The function $w(d)$ shows only a weak dependence on $d$, like the relative width $n_{FWHM}/n_{MP}$ from Eq. \ref{mp_fwhm}. When going from a $50\,\mu$m to a $300\,\mu$m sensor this statistical effect improves only by 20-30\,\%. \\
Neglecting this weak dependence  on $d$, the time resolution at constant electric field i.e. at constant drift velocity $v_1$ and $v_2$ scales with $d$, which represents the trivial fact that the duration of the signal and therefore also $\Delta_{\tau}$ scales with $d$. For a given voltage $V$, the electric fields in the thinner sensors, and therefore the velocities of electrons and holes are of course larger, so the time resolution improves significantly beyond the $1/d$ scaling for thin sensors. 
\\
If we associate $v_1$ and $v_2$ with the electron and hole velocity, $T_1=d/v_1$ and $T_2=d/v_2$ are the total drift times of electrons and holes, and $T_{12}=d/\sqrt{v_1v_2}$ is the total drift time assuming the geometric mean of the electron and hole velocity, and the time resolution reads as
\beq \label{f_coefficients}
   \Delta_{\tau} =
        w(d)\, \sqrt{4/180} \, 
     \sqrt{T_1^2-1.75\,T_{12}^2+T_2^2}  
\eeq
To get realistic estimates we use an approximation for the velocity of the electrons and holes from \cite{canali}
\beq
     v_e(E) = 
     \frac{
         \mu_e\,E
     }{
         \left[
         1+\left(\frac{\mu_e\,E}{v^e_{sat}}\right)^{\beta_e}
         \right]^{1/\beta_e}    
     } 
     \qquad \qquad
     v_h(E) = 
     \frac{
         \mu_h\,E
     }{
         \left[
         1+\left(\frac{\mu_h\,E}{v^h_{sat}}\right)^{\beta_h}
         \right]^{1/\beta_h}    
     } 
\eeq
where we chose $\mu_e=1417$\,cm$^2$/Vs, $\mu_h=471$\,cm$^2$/Vs, $\beta_e=1.109$, $\beta_h=1.213$ and $v^e_{sat}=1.07\times10^7$\,cm/s and $v^h_{sat}=0.837\times10^7$\,cm/s at 300\,K in accordance with the default models in Sentaurus Device \cite{sentaurus}. The resulting drift velocity  together with the time that the electrons and holes need to traverse the sensor (assuming $V_{dep}=0$) are given in Fig. \ref{velocity}. For a 50\,$\mu$m sensor at 200\,V the electrons take 0.6\,ns and the holes take 0.8\,ns to traverse the sensor, so the total signal duration is $<0.8$\,ns.  \\
The values for the time resolution according to Eq. \ref{final_formula} for the Landau theory, the PAI model and a Gaussian fit to the PAI model are given in Fig. \ref{reso} for 50, 100, 200 and 300\,$\mu$m sensors. A 200\,$\mu$m sensor can achieve a time resolution of $<50\,$\,ps for $V>350$\,V and a 50\,$\mu$m sensor can achieve $<15$\,ps for $V>200$\,V.


\subsection{Multiple particles passing a silicon sensor}

In \cite{Akchurin2017} the time resolution for multiple particles crossing a sensor is discussed. The case of $n$ particles passing the silicon sensor is equivalent to the situation of one particle passing the sensor with a mean free path between collisions reduced to to $\lambda_n=\lambda/n$. According to the Landau theory we have $w(d) \approx 1/\sqrt{\ln d/\lambda}$ for a single particle, so for $n$ particles the fluctuations reduce according to 
\beq \label{multiple_particle_formula}
   \frac{\Delta_{\tau}(n\, \mbox{particles})}{\Delta_{\tau}(1\, \mbox{particle})} = 
   \frac{1}{\sqrt{1+\frac{\ln n}{\ln d/\lambda}}} 
\eeq 
This function has an extremely weak dependence on $n$ so the improvement of the centroid time resolution when going from $1$ to $100$ particles for a 50/100/200/300\,$\mu$m sensor is only 26/24/23/22\,\%. The centroid  time resolution does therefore not significantly change for multiple particles. The signal to noise ratio does however improve almost linearly with the number of particles passing the sensor, so when using leading edge discrimination with a threshold set close to the noise level as discussed in Section \ref{leading_edge_section}, there is in principle no lower bound on the time resolution.

%%%%%%%%%%%%%%%%%%%%%%%%%%%%%%%%%%%%%
%%%%%%%%%%%%%%%%%%%%%%%%%%%%%%%%%%%%%
%%%%%%%%%%%%%%%%%%%%%%%%%%%%%%%%%%%%%
%%%%%%%%%%%%%%%%%%%%%%%%%%%%%%%%%%%%%
%%%%%%%%%%%%%%%%%%%%%%%%%%%%%%%%%%%%%
%%%%%%%%%%%%%%%%%%%%%%%%%%%%%%%%%%%%%

\subsection{Noise contribution to the centroid  time resolution}

\begin{figure}
 \begin{center}
 a)
  \includegraphics[width=7cm]{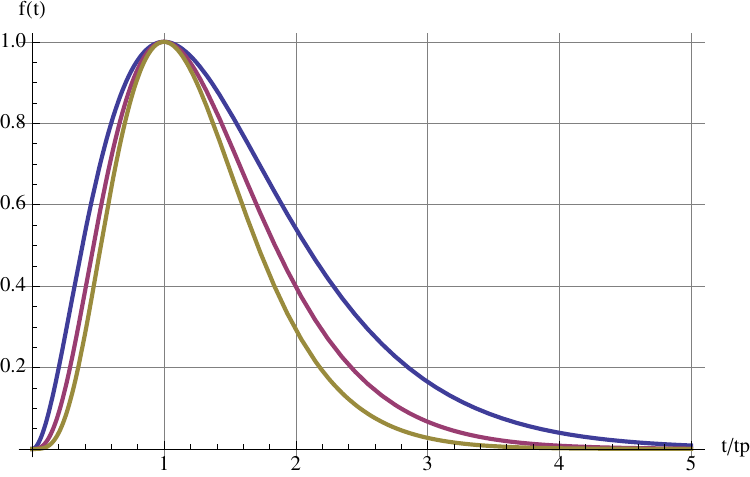}
  b)
  \includegraphics[width=7cm]{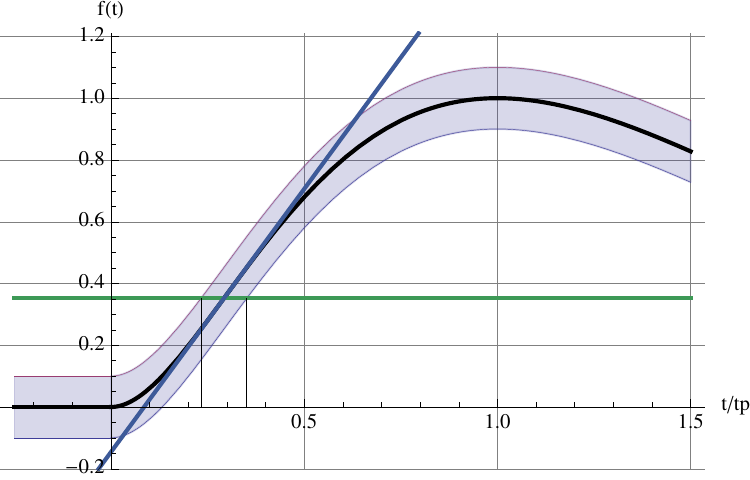}
  \caption{ a) Amplifer response for $n=2,3,4$ from Eq. \ref{delta_response}. b) Contribution to the time resolution from the noise. }
  \label{delta_reponse_plot}
  \end{center}
\end{figure}

\begin{figure}
 \begin{center}
   \includegraphics[width=10cm]{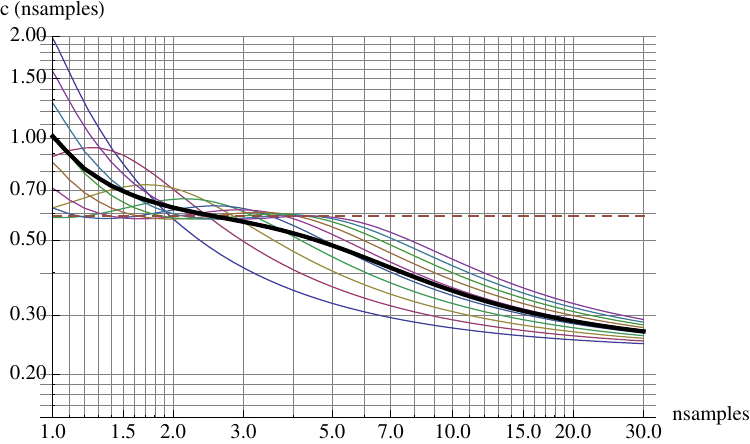}
  \includegraphics[width=10cm]{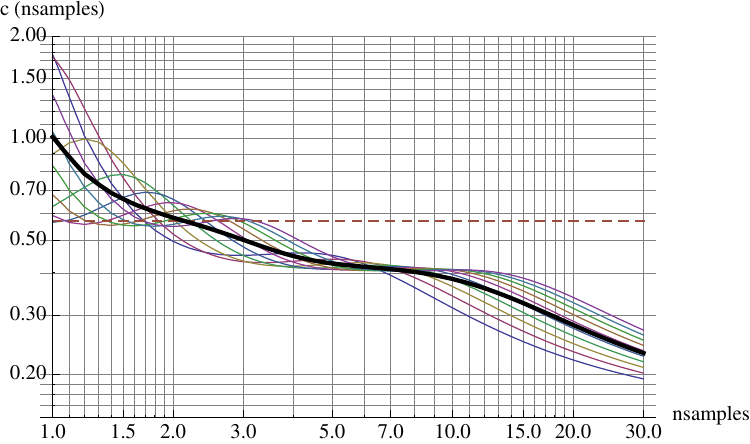}
  \caption{The function $c(n_{s})$ for an amplifier with $n=2$ (top) and $n=3$ (bottom). The horizontal line is the result for constant fraction discrimination at the maximum slope from Eq. \ref{average_noise}. }
  \label{fitplot}
  \end{center}
\end{figure}

As shown in Eq. \ref{cog_delta} the centroid  time of a signal can be measured by using an amplifier with a peaking time $t_p$ that is larger than the total signal time $T$. For a 50\,$\mu$m sensor at 250\,V this signal time is $T \approx 0.8$\,ns, so an amplifier with peaking time $t_p>1.5$\,ns can realise such a measurement. The problem to solve is therefore to measure the time of a pulse of known shape (the delta response) that has noise of a known frequency spectrum superimposed. This can be accomplished by various techniques of constant fraction discrimination or continuous sampling with optimum filtering methods, both of which will be discussed in this section. For the remainder of the report we assume an unipolar amplifier with a delta response of
\beq \label{delta_response}
  f(t)=\left( \frac{t}{t_p}\right)^n\,e^{n(1-t/t_p)}\,\Theta(t)
\eeq
where $t_p$ is the peaking time and $\Theta(t)$ is the Heaviside step function. The delta response for $n=2, 3, 4$ is shown in Fig. \ref{delta_reponse_plot}a. Such an amplifier can be realized by $n$ integration stages with $\tau=RC=t_p/n$ and for large values of $n$ it approaches Gaussian shape (semi-gaussian shaping). In general we can use it to parametrize a measured delta response shape by adjusting $n$ and $t_p$ to fit a specific amplifier delta response. The normalized transfer function and related 3\,dB bandwidth frequency $f_{bw}$ of the above delta response are given by
\beq
   \vert W(i2\pi f)\vert = \frac{1}{\sqrt{[1+(2\pi f)^2t_p^2/n^2]^{n+1}}} 
   \qquad \qquad
   f_{bw}=\frac{1}{2\pi\,t_p} \,n\sqrt{2^{1/(n+1)}-1}
\eeq
For constant fraction discrimination we set the threshold to a value where $f(t)$ has the maximum slope of $f'(t_s)$ at time $t_s$ which evaluates to  
\beq
  t_s=t_p\,(1-1/\sqrt{n})\qquad f'(t_s)=\frac{1}{t_p}\,e^{\sqrt{n}}n^{(3/2-n)}(n-\sqrt{n})^{n-1}
\eeq
Assuming a pulse-height $A$ and a noise of $\sigma_{noise}$, the timing error when applying the threshold at the maximum slope is then 
\beq
   \sigma_t= \frac{\sigma_{noise}}{A}\,\frac{1}{f'(t_s)} = 
   \frac{\sigma_{noise}}{A} \, \frac{t_p}{\,e^{\sqrt{n}}n^{(3/2-n)}(n-\sqrt{n})^{n-1}} =
   \frac{\sigma_{noise}}{A} \, \frac{1}{2\pi f_{bw}}\,\frac{   \sqrt{2^{1/(n+1)}-1} }{\,e^{\sqrt{n}}n^{(1/2-n)}(n-\sqrt{n})^{n-1}}
\eeq
as illustrated in Fig. \ref{delta_reponse_plot}b. This evaluates to 
\begin{eqnarray} \label{cf_time_resolution}
  &=& \frac{\sigma_{noise}}{A}\,\,\,\,t_p \,\,\times (0.59, 0.57, 0.54, 0.51) \qquad \mbox{for} \qquad n=2,3,4,5 \\
  & = &  \frac{\sigma_{noise}}{A} \, \frac{1}{f_{bw}} \times (0.10, 0.12, 0.13, 0.14) \qquad \mbox{for} \qquad n=2,3,4,5 \no
\end{eqnarray}
So for an amplifier with a peaking time of $t_p$=1\,ns and $n=2$, the time resolution is 60\,ps for a signal to noise ratio of 10 and 20\,ps for a signal to noise ratio of 30. \\
The pulse-height of the silicon sensor signal is given by the total number $n$ of deposited e-h pairs, so if we write the noise $\sigma_{noise}$ in units of electrons, the average expression for $\sigma_{noise}/A$ becomes 
\beq
 \frac{\sigma_{noise}}{A}= \int_0^\infty \frac{\sigma_{noise}}{n}\,  p(n, d)\,dn 
\eeq
where $p(n, d)$ is from Eq. \ref{general_pn_distribution}. For the Landau theory we use Eq. \ref{bN} to evaluate this expression to 
\beq
    \int_0^\infty \frac{\sigma_{noise}}{n}\,  p(n, d)\,dn = \frac{\sigma_{noise}\,\lambda}{n_0\,d}\,w_1(d)^2 
    \approx  \frac{\sigma_{noise}\,\lambda}{n_0\,d}\,\frac{1}{a_1+b_1\ln d/\lambda}
\eeq
For the average time resolution we therefore find 
\begin{eqnarray}
    \ov \sigma_t 
   & \approx & \sigma_{noise}\frac{ \lambda}{n_0 \,d}\, \frac{1}{1+1.155\ln d/\lambda}\,\,\,\,t_p \,\,\times (0.59, 0.57, 0.54, 0.51) \qquad \mbox{for} \qquad n=2,3,4,5 \label{average_noise} \\
  & \approx & \sigma_{noise}\frac{\lambda}{n_0\,d}\, \frac{1}{1+1.155\ln d/\lambda}\, \frac{1}{f_{bw}} \times  (0.10, 0.12, 0.13, 0.14) \qquad \mbox{for} \qquad n=2,3,4,5
\end{eqnarray}
For an average cluster distance of $\lambda = 0.212\,\mu$m, $n_0=2.2$ and an amplifier with $n=2$,  this expression becomes 
\bea
   \sigma_t  & \approx & \sigma_{noise}[\mbox{electrons}] \, \times \,1.6\times 10^{-4} \, t_p \qquad d = 50\mu m \\
                   & \approx & \sigma_{noise}[\mbox{electrons}] \, \times \,3.3\times 10^{-5} \, t_p \qquad d = 200\mu m 
\eea
Assuming a 50\,$\mu$m sensor and a peaking time of 2\,ns and an Equivalent Noise Charge (ENC) of 50 electrons, the noise contribution to the time resolution is 16.6\,ps. Assuming a 200\,$\mu$m sensor and $t_p=10$\,ns and and ENC of 200 electrons, the contribution to the time resolution is 66\,ps. The series noise of an amplifier for a given white series noise spectral density $e_n^2$ and detector capacitance $C$ is given by 
\beq
  \sigma_{noise}^2=  \frac{1}{2}e_n^2C^2\int_{-\infty}^\infty f'(t)^2dt = 
  \frac{1}{2}e_n^2 \, C^2 
  \frac{n^2\,(2n-2)!}{t_p}
   \left(\frac{e}{2n}
   \right)^{2n}
\eeq
For constant $e_n^2$ the noise decreases with $1/\sqrt{t_p}$ while the time resolution is proportional to $t_p$, so one favours short peaking times for minimizing the impact of noise, as long as other noise sources do not become dominant.
\begin{figure}
 \begin{center}
 a)
  \includegraphics[width=7cm]{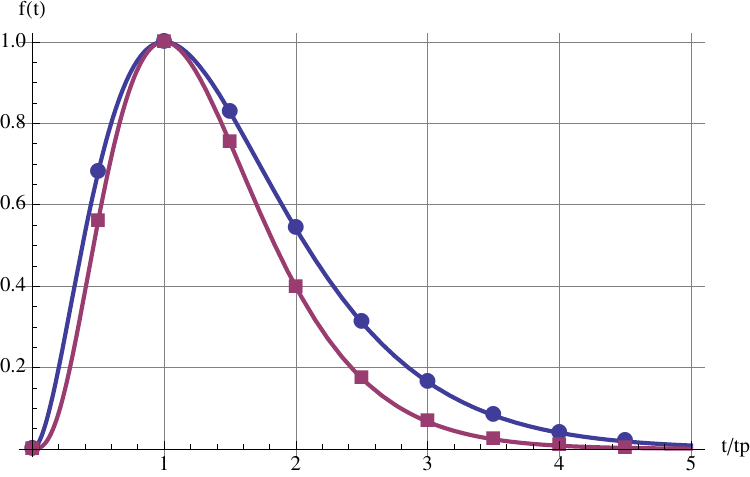}
b)
  \includegraphics[width=7cm]{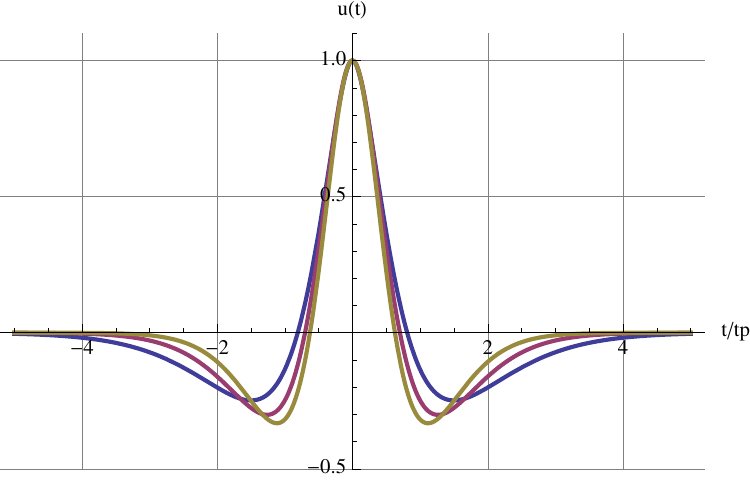}
  \caption{a) Sampling the signal at constant frequency. b) Autocorrelation function of $f'(t)$ for $n=2, 3, 4$. For times smaller than $0.5\,t_p$ the samples become highly correlated. }
  \label{sampling}
  \end{center}
\end{figure}
\\ \\
Since we know the shape of the delta response, continuous sampling of the signal and fitting of the known shape to the sample points provides an effective way to determine the time as shown in Fig. \ref{sampling}a and investigated in the following. We have to fit the function $A\,f(t-\tau)$ to the measured signal with the amplitude $A$ and time $\tau$ as free parameters. Linearizing this expression for small values of $\tau$ we have 
\beq
   A\,f(t-\tau) \approx A\,f(t)-A\,f'(t)\tau = \alpha_1 \, f(t)-\alpha_2\,f'(t)
   \qquad
   \alpha_1=A \qquad \alpha_2=A\tau
\eeq
Finding the best estimate of $\alpha_1, \alpha_2$ for a signal signal $S_1, S_2, ... , S_N$ sampled at times $t_1, t_2, ..., t_N$ leads to the familiar problem of linear regression. We proceed as outlined in \cite{optimum_filter} where the problem is stated as a $\chi^2$ minimization according to
\beq
  \chi^2 = \sum_{i=1}^N\sum_{j=1}^N 
  [S_i-\alpha_1f(t_i)+\alpha_2 f'(t_i)]\,
  V_{ij}
   [S_j-\alpha_1f(t_j)+\alpha_2 f'(t_j)]\, 
\eeq
The matrix $V_{ij}$ is the inverse of the autocorrelation matrix $R_{ij}=R(t_i-t_j)$ with $R(t)$ being the autocorrelation function of the noise. %
The autocorrelation function of this series noise is 
\beq
  R(t)= \sigma_{noise}^2 \,  \int_{-\infty}^\infty f'(t+u)\,f'(u)du = 
   \sigma_{noise}^2\,n!\,  \left(
    \frac{2n \vert t \vert}{t_p}
    \right)^n
    \frac{2t_pK_{n-1/2}(n\vert t \vert /t_p)-tK_{n+1/2}(n \vert t \vert/t_p)}{(2n-2)!\,
    \sqrt{2 n \vert t \vert \, t_p \pi}}
\eeq
with $K_{\nu}(x)$ being the modified Bessel function of the second kind. For $n=2, 3$ evaluates to
\bea
    R(t) = \sigma_{noise}^2 U(t)&=& \sigma_{noise}^2\,e^{-2\vert t \vert/t_p}\,\left[ 
    1+2\,\frac{\vert t \vert}{t_p}-4\left(\frac{\vert t \vert}{t_p}\right)^2  \right] \qquad n=2 \\
    &=& \sigma_{noise}^2\,e^{-3\vert t \vert /t_p}\,\left[ 
    1+3\,\frac{\vert t \vert}{t_p}-9\left(\frac{\vert t \vert}{t_p}\right)^3  \right] \qquad n=3 
\eea
The autocorrelation function is shown in Fig. \ref{sampling}b, and we see that for time intervals smaller than $t_p/2$ the samples become highly correlated. In the following we us $n_s$ samples within the peaking time $t_p$, so we have sampling time bins of $\Delta t=t_p/n_s$. We sample the signal in the range of $0<t<5\,t_p$, giving $t_i= i\,\Delta t$ with $0<i<5\,n_s$. Defining 
\beq
  Q_1(n_s) =\sum_{ij} f(t_i)U^{-1}_{ij}f(t_j) \quad 
  Q_2 (n_s)=\sum_{ij} f'(t_i)U^{-1}_{ij}f'(t_j) \quad 
  Q_3 (n_s)=\sum_{ij} f'(t_i)U^{-1}_{ij}f(t_j) 
\eeq
where $U^{-1}_{ij}$ is the inverse of the matrix $U_{ij}=U(t_i-t_j)$, the covariance matrix elements $\varepsilon_{ij}$ for $\alpha_1, \alpha_2$ are then 
\beq
   \varepsilon_{11}=\sigma_A^2 = \frac{\sigma_{noise}^2\,Q_2}{Q_1Q_2-Q_3^2} \qquad
   \varepsilon_{22}=A^2\,\frac{\sigma_\tau^2}{t_p^2} = \frac{\sigma_{noise}^2\,Q_1}{Q_1Q_2-Q_3^2} \qquad
   \varepsilon_{12}= \frac{\sigma_{noise}^2\,Q_3}{Q_1Q_2-Q_3^2} 
\eeq
So for the time resolution we finally have
\beq
     \frac{\sigma_\tau}{t_p} = \frac{\sigma_{noise}}{A}\,\sqrt{\frac{Q_1(n_s)}{Q_1(n_s)Q_2(n_s)-Q_3(n_s)^2}} = \frac{\sigma_{noise}}{A}\, c(n_s)
\eeq
Using as before the average signal to noise ratio for a sensor of thickness $d$ we find 
\beq
    \ov{\sigma_t}  =  \sigma_{noise} [\mbox{electrons}] \frac{\lambda}{n_0\,d}\, \frac{1}{1+1.155\ln d/\lambda}\,t_p \, c(n_s)
 \eeq
This expression represents the optimum time resolution that can be achieved for a given sampling frequency.
Fig. \ref{fitplot} shows the function $c(n_s)$ assuming an amplifier with $n=2, 3$. The horizontal lines correspond to the numbers of 0.59 and 0.57 from Eq. \ref{average_noise} when using constant fraction discrimination at the maximum slope.
The families of curves represent a scan of the sampling phase with respect to the peak of the signal and the solid curve represents the average. The samples on the largest slope carry the highest weight on time information, while samples around the signal peak carry very little time information. \\
We see that sampling at an interval corresponding to half the peaking time $(n_s=2)$ gives approximately the same result as the constant fraction discrimination at maximum slope. By increasing the sampling rate further the value cannot be improved much beyond a factor 2-3. This result is quite evident, since the noise is highly correlated on a timescale of $<t_p/2$ as seen from Fig. \ref{sampling}b, so further increase of the sampling rate does not provide more information. 

\clearpage

%%%%%%%%%%%%%%%%%%%%%%%%%%%%%%%%%%%%%%%%%
%%%%%%%%%%%%%%%%%%%%%%%%%%%%%%%%%%%%%%%%%
%%%%%%%%%%%%%%%%%%%%%%%%%%%%%%%%%%%%%%%%%
%%%%%%%%%%%%%%%%%%%%%%%%%%%%%%%%%%%%%%%%%
%%%%%%%%%%%%%%%%%%%%%%%%%%%%%%%%%%%%%%%%%

\subsection{Weighting field effect on the centroid  time for uniform charge deposit}

\begin{figure}
 \begin{center}
    a)
    \includegraphics[width=4cm]{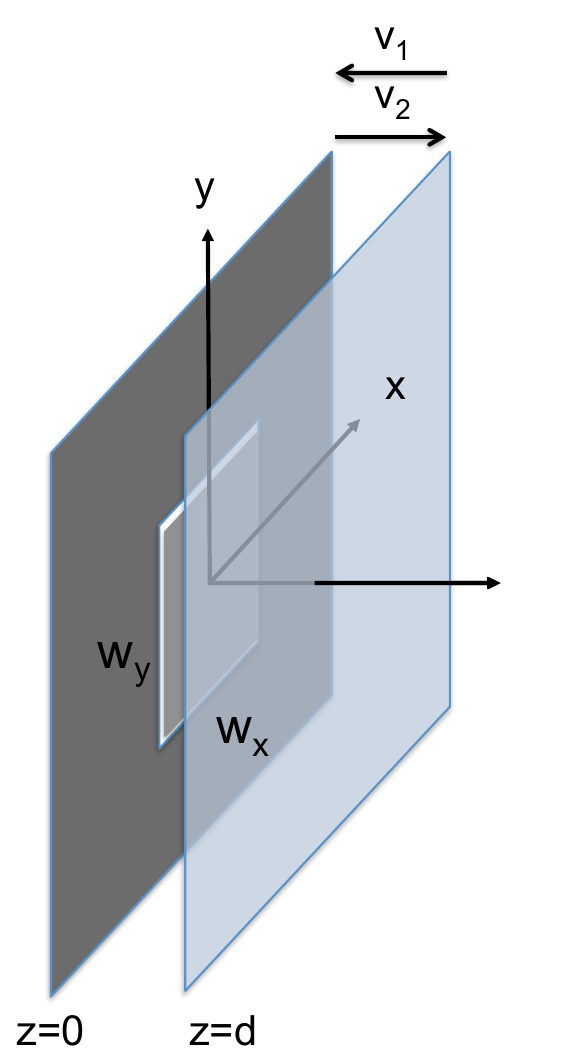}
    \hspace{2cm}
      \includegraphics[width=6cm]{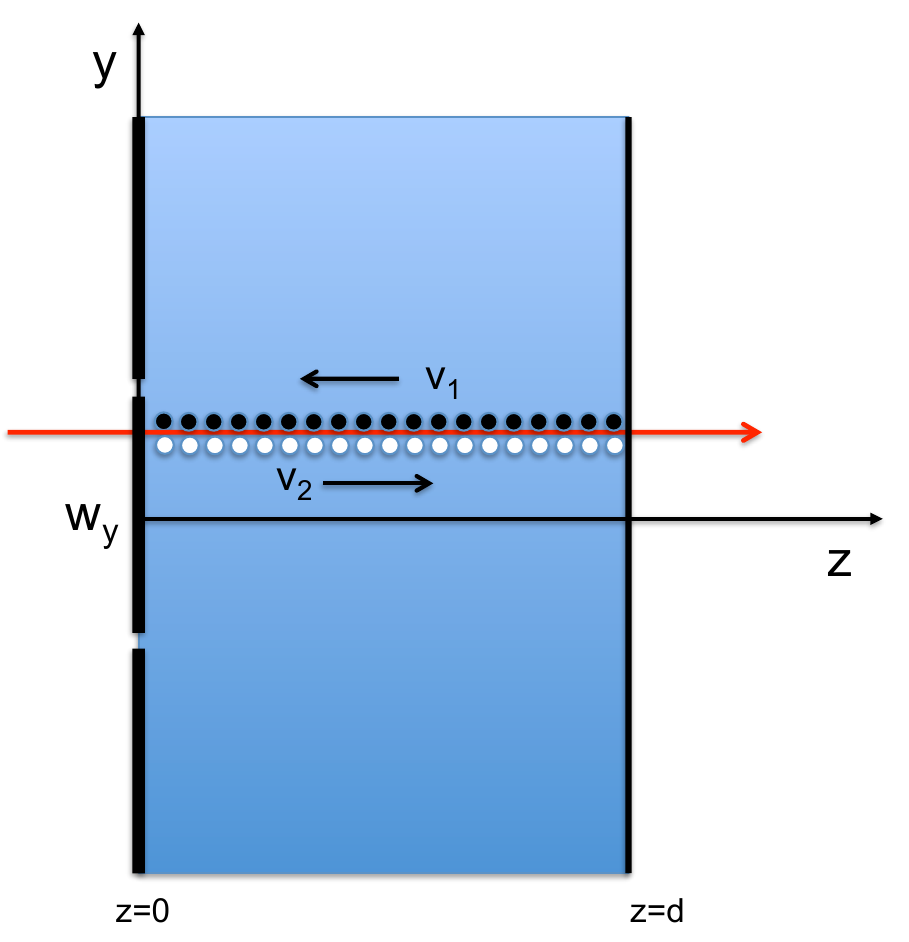}
      b)
  \caption{a) A pixel of dimension $w_x, w_y$ centred at $x=y=z=0$ in a parallel plate geometry of plate distance $d$. b) Uniform charge deposit of a particle passing the silicon sensor. $v_1$ is the velocity of charges moving towards the pixel and $v_2$ is the velocity of charges moving away from the pixel.}
  \label{pad_figure}
  \end{center}
\end{figure}

\begin{figure}
 \begin{center}
  a) \includegraphics[width=4.5cm]{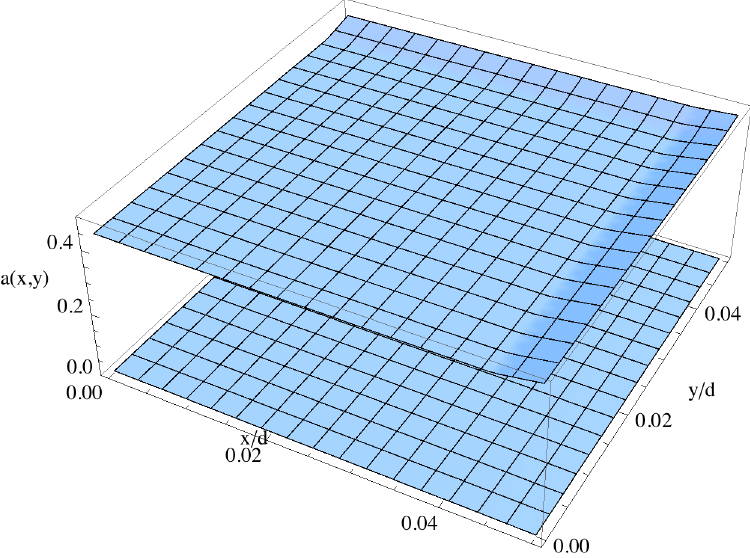}
   b) \includegraphics[width=4.5cm]{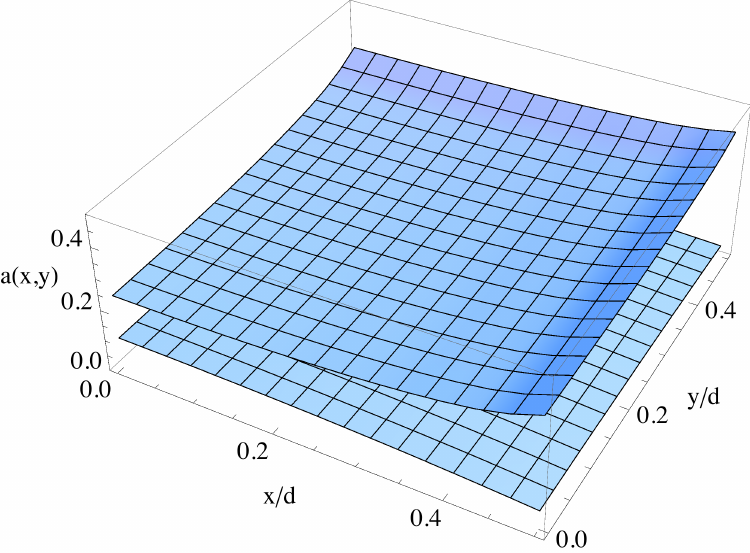}
   c) \includegraphics[width=4.5cm]{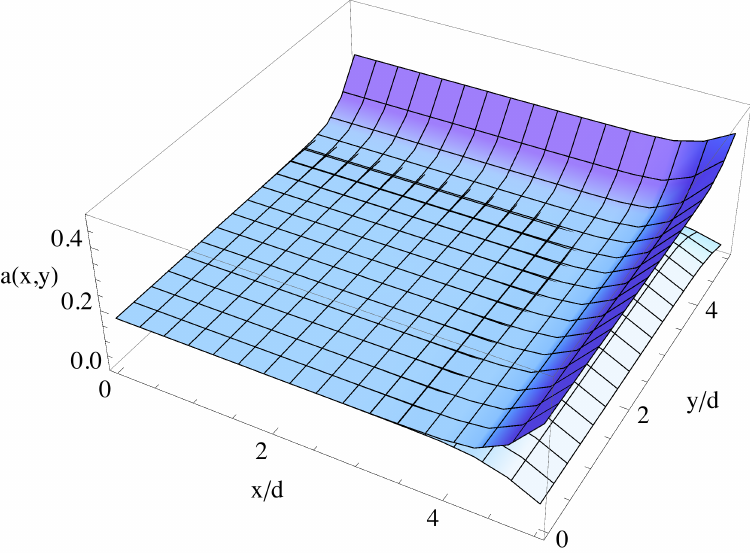}
  \caption{The functions $a_1(x, y)$ and $a_2(x, y)$ from Eq. \ref{tau_uniform} that determine the centroid  time for a signal from two line charges $-q_{line}, q_{line}$ at position $x, y,$. The top graph corresponds to $a_1$ and the bottom one to $a_2$. The three plots correspond to pads of size a) $w/d=0.1$, b) $w/d=1$, c) $w/d=10$.   }
  \label{a_functions}
  \end{center}
\end{figure}

\begin{figure}
 \begin{center}
  \includegraphics[width=7cm]{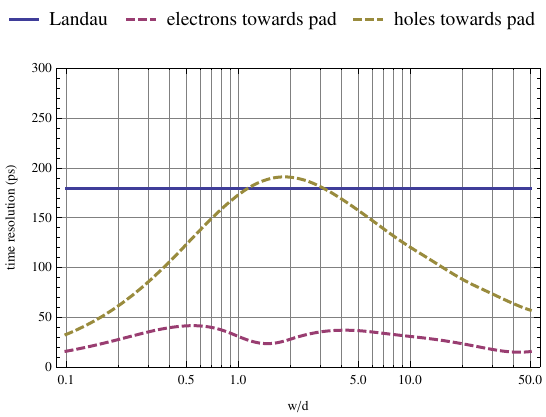}
  \includegraphics[width=7cm]{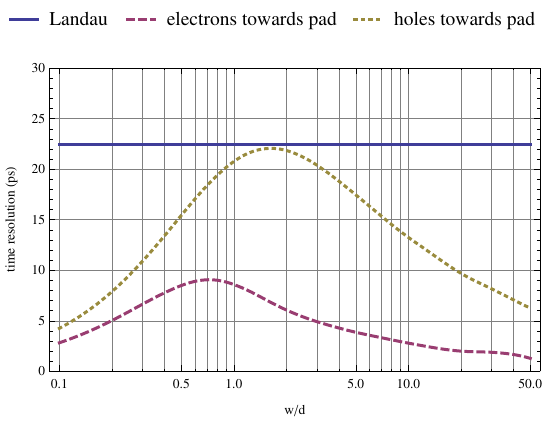}
  \caption{Standard deviation for the centroid  time for sensor thickness of a) $d=200\,\mu$m and b) $d=50\,\mu$m and $V=200$\,V, assuming uniform charge deposit and a square readout pad. The horizontal line represents centroid  time resolution from Eq. \ref{final_formula} due to Landau fluctuations only. The two curves in the plots represent the effect of weighting field fluctuations where either the electrons or the holes move towards the readout pad.
  }
  \label{uniform}
  \end{center}
\end{figure}

Up to now we have assumed the sensor readout electrode to be represented by an infinite parallel plate capacitor, which in practice corresponds to readout pads or pixels that are much larger than the sensor thickness $d$. In many practical applications, the granularity is however similar to the sensor thickness. The shape of the induced signal therefore becomes dependent on the $x, y$ position of the track and the centroid  time will be affected. In this section we investigate this effect by using the weighting field of a rectangular pixel as presented in \cite{weighting_field}, shown in Fig. \ref{pad_figure}a and detailed in \ref{appendix_weightfield}.\\
We assume again the sensor to be represented by a parallel plate geometry between $z=0$ and $z=d$ and assume charges to move along the z-axis. We also assume normal incidence of the particle and negligible diffusion. The plate at $z=0$ is segmented into pixels such that we find a weighting field of $E_w(x, y, z)=-d\phi_w(x, y, z)/dz$ along the z-axis. We first assume a single charge pair to be produced at position $x, y, z$ with $-q$ moving towards the the pixel at $z=0$ according to $z_1(t)=z-v_1t$ and $+q$ moving towards the plate at $z=d$ according $z_2(t)=z+v_2t$, so the induced current becomes  
\bea
     \frac{i(t)}{q} & = & E_w[x,y,z_1(t)]\dot{z}_1(t)\,\Theta(z/v_1-t) +
    E_w[x,y,z_2(t)]\dot{z}_2(t)\,\Theta((d-z)/v_2-t) \\
     & = & - v_1  E_w[x,y,z-v_1t]\Theta (z/v_1-t)
     - v_2 E_w[x,y,z+v_2t]\Theta ((d-z)/v_2-t)
\eea
The centroid  time of this signal is
\beq \label{single_charge_tau}
    \tau(x,y,z) = \frac{\int t \,i(t) dt}{\int i(t)dt}=\frac{d}{v_1}\, \Psi_1(x,y,z) +  \frac{d}{v_2}\, \Psi_2(x,y,z)
\eeq
\beq
  \Psi_1(x,y,z) =   \frac{z}{d}-\frac{1}{d}\int_0^z \phi_w(x,y,z')dz'
  \qquad
  \Psi_2(x,y,z) =  \frac{1}{d}\int_z^d \phi_w(x,y,z')dz'
\eeq
In case there is not a single pair of charges $q, -q$ but a pair of uniform line charges between $z=0$ and $z=d$, as shown in Fig. \ref{pad_figure}b, we have
\bea
     \frac{I(x,y,t)}{q_{line}}  & = & -v_1  \int_0^d E_w[x,y,z-v_1t]\Theta (z/v_1-t)dz
     -v_2 \int_0^d E_w[x,y,z+v_2t]\Theta ((d-z)/v_2-t)dz  \no  \\
     & = & -v_1  \left[
     1-\phi_w(x,y,d-v_1t)
     \right]  \Theta(d/v_1-t)
     - v_2\, \phi_w(x,y,v_2 t) \, \Theta(d/v_2-t)
\eea
where $q_{line}$ is the charge per unit of length. The centroid  time of this signal then reads as
\beq \label{tau_uniform}
   \tau (x,y) = \frac{d}{v_1}\,a_1(x,y) + \frac{d}{v_2}\,a_2(x,y) = 
  T_1\,a_1(x,y) +T_2\,a_2(x,y)
\eeq
\bea
  a_1(x,y) &=& \frac{1}{d} \int_0^d \Psi_1(x,y,z) dz = 
   \frac{1}{2}-\frac{1}{d^2} \int_0^d(d-z)\phi_w(x,y,z)dz   \\
  a_2(x,y) &=& \frac{1}{d} \int_0^d \Psi_2(x,y,z) dz = 
   \frac{1}{d^2} \int_0^dz\phi_w(x,y,z)dz   
\eea
The two functions $a_1(x,y)$ and $a_2(x,y)$ are shown in Fig. \ref{a_functions}. We can see that for large pads the values for both functions approach the constant value of $1/6$ in accordance with Eq. \ref{tau_average_parallel_plate} with some deviations at the border. For small pads the average of $a_1$ and $a_2$ is quite different, but the functions are also quite uniform. For the pad size of $w/d \approx 1$ the two functions vary significantly across the pad, which we will quantify next.
In case the pixel is uniformly irradiated, the probability to hit an area $dx\,dy$ is given by $dx\,dy/(w_x w_y)$ and the average centroid  time, the second moment and the standard deviation $\Delta _{\tau}$ are given by
\beq
   \ov \tau = \frac{1}{w_x\,w_y} 
   \int_{-w_x/2}^{w_x/2}
   \int_{-w_y/2}^{w_y/2}
   \tau(x,y)
   dx dy
   \qquad
      \ov {\tau^2} = \frac{1}{w_x\,w_y} 
   \int_{-w_x/2}^{w_x/2}
   \int_{-w_y/2}^{w_y/2}
   \tau^2(x,y)
   dx dy
\eeq
\beq \label{weighting_field_resolution1}
  \Delta_\tau^2=\ov{\tau ^2} - 
  \ov \tau^2 = 
 d^2 \, \left( 
 \frac{c_{11}}{v_1^2} +
 \frac{c_{12}}{v_1 v_2} +
 \frac{c_{22}}{v_2^2}
  \right)
  = c_{11}T_1^2+c_{12}T_{12}+c_{22}T_2^2
\eeq
where we have defined 
\bea
       c_{11} & = &    \frac{1}{w_x\,w_y}
   \iint a_1^2 dx dy
   -  \left( \frac{1}{w_x\,w_y}\iint a_1 dx dy\right)^2  \label{c11} \\
       c_{12} & = &   \frac{2}{w_x\,w_y} 
   \iint  a_1a_2 dx dy
   -  \frac{2}{(w_x\,w_y)^2}\iint  a_1 dx dy\,
   \iint  a_2 dx dy  \label{c12} \\
    c_{22} & = &    \frac{1}{w_x\,w_y} 
   \iint a_2^2 dx dy
   - \left( \frac{1}{w_x\,w_y}\iint  a_2 dx dy\right)^2 \label{c22}
\eea
and
\beq
    T_1=d/v_1 \qquad T_2 =d/v_2 \qquad T_{12} = d/\sqrt{v_1 v_2}
\eeq
Before moving to the numerical evaluation we investigate the limiting cases for very large and very small pads.
For large pixels we have $\phi_w=1-z/d$ and the expressions become
\beq
    a_1(x,y) =  \frac{1}{6} \qquad 
    a_2 (x,y) = \frac{1}{6} \qquad 
    \mbox{for} \qquad
    w/d \gg 1
\eeq
which results in $\ov \tau = d/6(1/v_1+1/v_2)$ in accordance with Eq. \ref{tau_average_parallel_plate} for an infinite electrode. Since there is no dependence on $x$ and $y$, the coefficients $c_{11}, c_{12}, c_{22}$ vanish, which is the expected result for an infinitely electrode.
 \\ \\
For very small pads the weighting potential falls to zero very quickly as a function of $z$, from it's value of unity on the pad surface at $z=0$. The integrals of the weighting potential over $z$ will therefore vanish and we have
\beq
    a_1(x,y) =  \frac{1}{2} \qquad 
    a_2 (x,y) = 0 \qquad
     \mbox{for} \qquad
    w/d \ll 1
\eeq
For this case only the charges moving towards the pad with $v_1$ contribute to the centroid  time and the average centroid  time becomes $\ov \tau = d/2v_1$. Since the weighting potential and weighting field are concentrated around the pixel surface the charges that never enter this area, i.e. the charges moving with $v_2$ towards $z=d$ will not contribute to the signal. The coefficients $c_{11}, c_{12}, c_{22}$ will again vanish because $a_1$ and $a_2$ have no dependence on $x,y$. Because the two limiting cases are zero, this means that there will be a pad size where the effect of the weighting field fluctuation is maximal, as discussed in the following.
\\ \\
The numerical evaluation of Eqs. \ref{c11}, \ref{c12}, \ref{c22} for square pixels of width $w$ for different ratios of $w/d$ are given in Table \ref{ctable} of the Appendix and the graphical representation of the coefficients is shown in Fig. \ref{c_coefficients}. The weighting potential of a pixel as given in Eq. \ref{weighting_potential} of  the Appendix is used.
The weighting field effect on the time resolution is worst for pad sizes corresponding to about 2-3 times the sensor thickness $d$, where the $c_{11}$ and $c_{12}$ coefficients assume a value around $2 \times 10^{-3}$. The coefficient $c_{11}$ is related to $v_1$ i.e. to the charges moving to the readout pad, $c_{22}$ is related to the charges moving in opposite direction. Since $c_{11} > c_{22}$ by a significant factor, the time resolution will be better if $v_1>v_2$ i.e. if the electrons are moving towards the pixels. The contribution to the time resolution from Eq. \ref{weighting_field_resolution1} is shown in Fig. \ref{uniform}. In case the holes move towards the pixel we find a maximum for values of $w/d \approx 2$, where the contribution becomes similar to the value from Landau fluctuations. In case the electrons move towards the pixel, the contribution is significantly smaller with maxima around $w/d \approx 1$. \\
The somewhat slow decrease of the effect for pad sizes of $w/d>3$ is due to the fact that we are calculating the standard deviation of the centroid time. As shown in Fig. \ref{a_functions}c) for $w/d=10$ there is no variation of the centroid time in the central 70\% of the pixel area and the variations take place only at the edges. The resulting time distribution for uniform illumination is significantly non-Gaussian with long tails. The true impact on the time resolution therefore depends also on the method of using the measured time and the algorithm for defining the time resolution.
\\
The final resolution is not given by the square sum of the Landau fluctuations from Eq. \ref{final_formula} and the weighting field fluctuations from Eq. \ref{weighting_field_resolution1}, since there is a very strong correlation between the two. This will be discussed in the next section.
\begin{figure}
 \begin{center}
  \includegraphics[width=12cm]{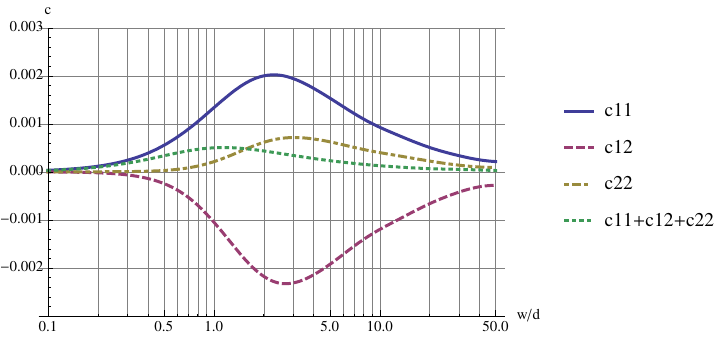}
  \caption{The coefficients $c_{11}, c_{12}, c_{22}$ for different values of $w/d$, where $w$ is the width of the square pad and $d$ is the silicon sensor thickness.}
  \label{c_coefficients}
  \end{center}
\end{figure}

\clearpage

%%%%%%%%%%%%%%%%%%%%%%%%%%%%%%%%%%%%%%%%%%%%%%
%%%%%%%%%%%%%%%%%%%%%%%%%%%%%%%%%%%%%%%%%%%%%%
%%%%%%%%%%%%%%%%%%%%%%%%%%%%%%%%%%%%%%%%%%%%%%
%%%%%%%%%%%%%%%%%%%%%%%%%%%%%%%%%%%%%%%%%%%%%%
%%%%%%%%%%%%%%%%%%%%%%%%%%%%%%%%%%%%%%%%%%%%%%
%%%%%%%%%%%%%%%%%%%%%%%%%%%%%%%%%%%%%%%%%%%%%%

\subsection{Centroid  time resolution for combined charge fluctuations and weighting field fluctuations}

\begin{figure}
 \begin{center}
  \includegraphics[width=8cm]{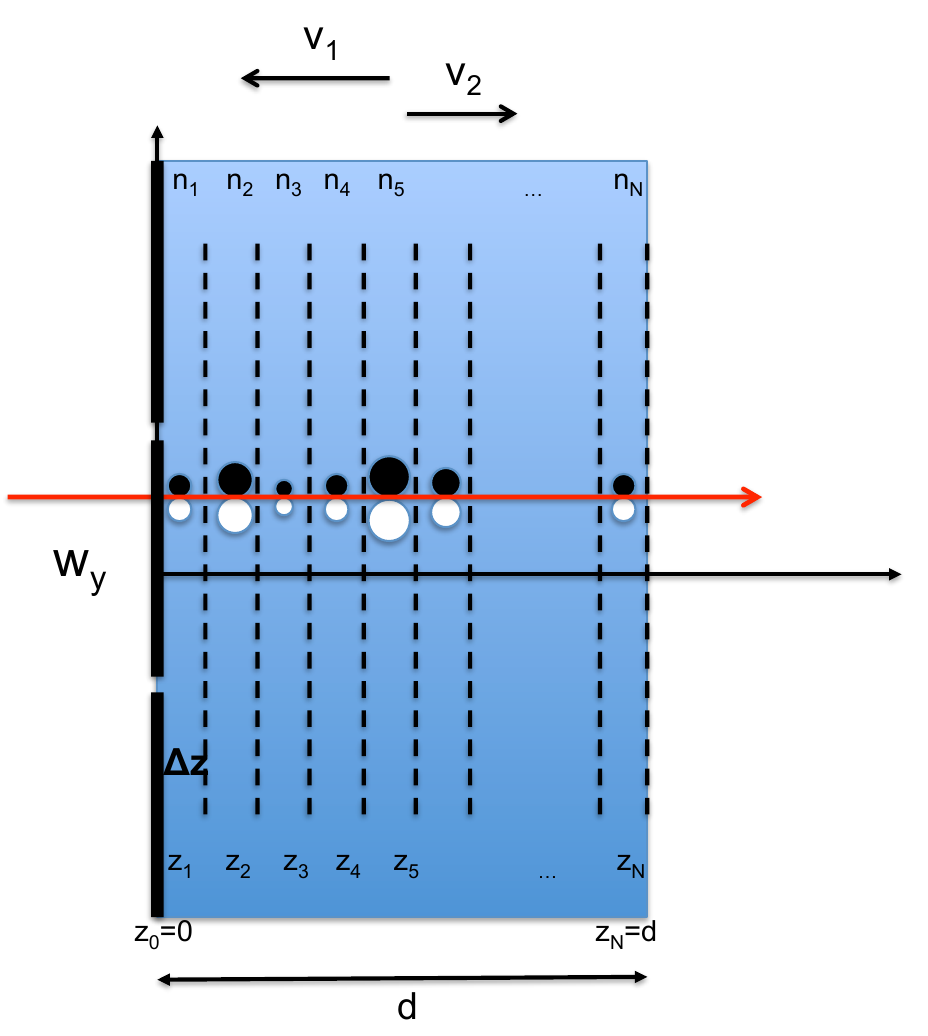}
  \caption{Silicon sensor with a readout pad centered at $x=y=z=0$. $v_1$ is the velocity of charges moving towards the pixel and $v_2$ is the velocity of charges moving away from the pixel.  }
  \label{sensor_slices_landau}
  \end{center}
\end{figure}
\begin{figure}
 \begin{center}
   a)
    \includegraphics[width=6.7cm]{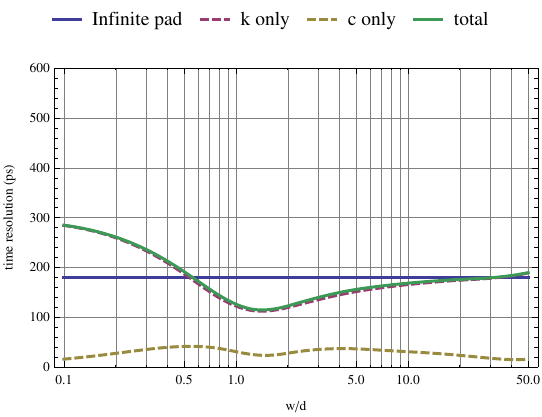}
    b)
    \includegraphics[width=6.7cm]{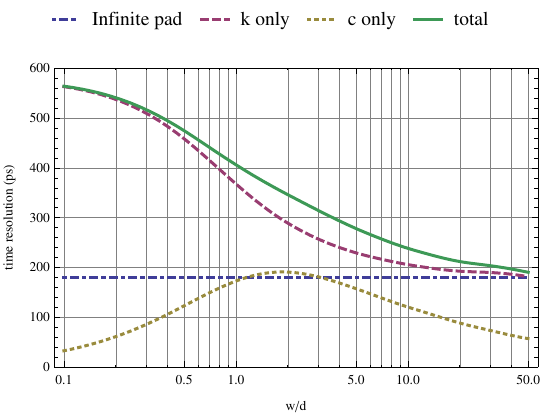}
  \caption{Centroid  time resolution for values of $d=200\,\mu$m and $V=200$\,V as a function of the pixel size $w$ assuming the Landau theory for the charge deposit. The 'c only' curve refers to the effect from a uniform line charge. In a) the electrons move towards the pixel while in b) the holes move towards the pixel.}
  \label{final_reso_200um}
  \end{center}
\end{figure}
\begin{figure}
 \begin{center}
 a)
    \includegraphics[width=6.7cm]{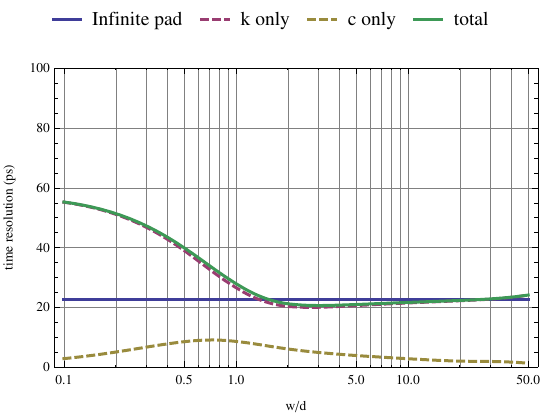}
    b)
    \includegraphics[width=6.7cm]{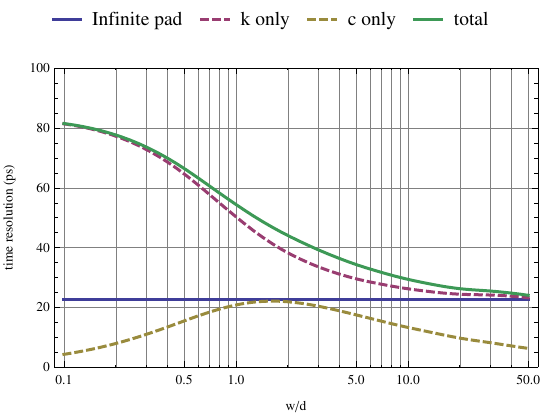}
  \caption{Time resolution for values of $d=50\,\mu$m and $V=200$\,V as a function of the pixel size $w$ assuming the Landau theory for the charge deposit. The 'c only' curve refers to the effect from a uniform line charge. In a) the electrons move towards the pixel while in b) the holes move towards the pixel.}
  \label{final_reso_50um}
  \end{center}
\end{figure}

In this section we consider the Landau fluctuations together with the variation of the $x,y$ position of the particle trajectory and the related fluctuation of the weighting field.  The centroid time for a particle that passes the sensor at position $x, y$ and deposits $n_k$ charges in the $N$ detector slices is given by
\beq
   \tau(n_1, n_2, ..., n_N, x, y)= 
   \frac{1}{\sum_{k=1}^N n_k}\, \sum_{k=1}^N n_k 
  \tau(x,y,k\Delta z)
\eeq
where $\tau(x,y,z)$ is from Eq. \ref{single_charge_tau}. Proceeding as detailed in \ref{appendix_statistics}  we calculate $\ov \tau$ and $\ov {\tau^2}$, where in addition to the integrals over $dn_1, dn_2, ..., dn_N$ we have to perform the integral $1/(w_xw_y) \int \int \tau dxdy$ for uniform illumination of a pad, and the final result for the variance is
\bea \label{variance_combined}
   \ov {\tau^2}-\ov {\tau}^2 &= &
  w(d)^2 \frac{1}{w_x w_y} 
    \iint \left[
     \frac{1}{d} \int_0^d\tau (x,y,z)^2dz -
     \left(
     \frac{1}{d} \int_0^d\tau (x,y,z)dz
     \right)^2
    \right]dx dy   \\
    & + & 
    \frac{1}{w_x w_y} \iint  \left(  \frac{1}{d} \int_0^d\tau (x,y,z) dz
     \right)^2 dx dy -
     \left[
     \frac{1}{w_xw_y} \iint  \left(\frac{1}{d} \int_0^d \tau (x,y,z) dz \right)dx dy
     \right]^2  \no
\eea
The second line of the expression is equivalent to the one considering the weighting field effect without charge fluctuations from the previous section, so the result can be expressed in the following terms
\beq \label{weighting_field_resolutionk}
    \Delta_{\tau}^2  = 
  w(d)^2 \, 
      \, \left( 
 \frac{k_{11}d^2}{v_1^2} +
 \frac{k_{12}d^2}{v_1 v_2} +
 \frac{k_{22}d^2}{v_2^2}
  \right)
 +
      \, \left( 
 \frac{c_{11}d^2}{v_1^2} +
 \frac{c_{12}d^2}{v_1 v_2} +
 \frac{c_{22}d^2}{v_2^2}
  \right) 
\eeq
The coefficients $c_{11}, c_{12}, c_{22}$ are the ones from the previous section and the coefficients $k_{11}, k_{12}, k_{22}$ are given by
\beq
   k_{11}=\frac{1}{w_x w_y} \iint (b_{11}-a_1^2)dx dy \qquad
   k_{12}=\frac{2}{w_x w_y} \iint (b_{12}-a_1a_2)dx dy \qquad
   k_{22}=\frac{1}{w_x w_y} \iint (b_{22}-a_2^2)dx dy \qquad
\eeq
with
\bea
    b_{11}(x,y) &=& \frac{1}{d} \int_0^d \Psi_1(x,y,z)^2 dz 
    = \frac{1}{d} \int_0^d   \left[  \frac{z}{d}-\frac{1}{d}\int_0^z \phi_w(x,y,z')dz'   \right]^2  dz \\
    b_{12}(x,y) &=& \frac{1}{d} \int_0^d \Psi_1(x,y,z)\Psi_2(x,y,z) dz 
            = \frac{1}{d} \int_0^d   \left[  \frac{z}{d}-\frac{1}{d}\int_0^z \phi_w(x,y,z')dz'   \right]
              \left[ \frac{1}{d}\int_z^d \phi_w(x,y,z')dz'\right]\,dz   \no   \\
    b_{22}(x,y) &=& \frac{1}{d} \int_0^d \Psi_2(x,y,z)^2 dz 
      = \frac{1}{d} \int_0^d   \left[ \frac{1}{d}\int_z^d \phi_w(x,y,z')dz'\right]^2  dz \no
\eea
First we verify the limiting cases for very large pads and very small pads. For large pads we substitute for the weighting potential the expression $\phi_w(x,y,z)=1-z/d$ and find
\beq
   b_{11}(x,y) =  \frac{1}{20}       \qquad
   b_{12}(x,y) =   \frac{1}{120}       \qquad
   b_{22}(x,y)=     \frac{1}{20}      \qquad
   w/d \gg 1
\eeq
which gives $k_{11}=k_{22}=4/180, k_{12}=-7/180$ and $c_{11}=c_{12}=c_{22}=0$, so we recuperate Eq. \ref{final_variance}. For very small pads the integrals of the weighting potential over $z$ will again vanish as discussed before, and we have
 \beq
   b_{11}(x,y) =  \frac{1}{3}     \qquad
   b_{12}(x,y) =  0   \qquad
   b_{22}(x,y)=    0    \qquad
   w/d \ll 1
\eeq
which gives $k_{11}=1/12, k_{12}=k_{22}=0$ and $c_{11}=c_{12}=c_{22}=0$ and therefore have 
\beq \label{arrival_time_formula1}
  \Delta_{\tau}= 
  w(d)\, 
      \frac{T_1}{\sqrt{12}}
\eeq
For small pads the weighting potential decays very quickly as a function of $z$, from its value of 1 on the pad surface to zero. The weighting field, which defines the induced current, is therefore very large close to the pad and zero for larger values of $z$. Only when the charges arrive at this position they will induce a signal. In the limiting case this is equivalent to a delta current signal for each charge that arrives at $z=0$, and we have
\beq \label{arrival_time_formula2}
  i(t) = q \sum_{k=1}^N n_k\, \delta(t-k \Delta z/v_1)  \qquad
   \tau = \frac{1}{\sum_{k=1}^N n_k} \, \sum_{k=1}^N n_k\,k\Delta z/v_1 \qquad
    \Delta_{\tau}= 
  w(d)\, 
      \frac{T_1}{\sqrt{12}}
\eeq
so we indeed recuperate the above expression for $\Delta_\tau$ ! We'll see the same formula later in Eq. \ref{gain_resolution} for silicon sensors with gain.  \\
The coefficients $k_{11}, k_{12}, k_{22}$ for square pads are listed in Table \ref{kktable} of the Appendix and are shown in Fig. \ref{k_coefficients}. The factor $k_{11}$, related to the charges moving with $v_1$ towards the pixel, is again larger than $k_{22}$, so as stated before the resolution is better if the electrons move towards the pixel. This fact is illustrated in Fig. \ref{final_reso_200um} and Fig. \ref{final_reso_50um} for a 200\,$\mu$m and 50\,$\mu$m sensor. It shows a significant difference for these two scenarios.  In case the electrons move to the pixel the weighting field effect seems not to add significantly to the time resolution for values of $w/d \gtrsim 1$.
\\
For pads with $w/d > 20$ one approaches the scenario of an infinitely extended electrode, as expected. For smaller pixels the Landau fluctuations and weighting field effect are strongly correlated and the resolution is significantly worse than expected from the quadratic sum of the weighting field effect for uniform charge deposit and the Landau fluctuation effects assuming an infinitely large electrode. 
\begin{figure}
 \begin{center}
  \includegraphics[width=10cm]{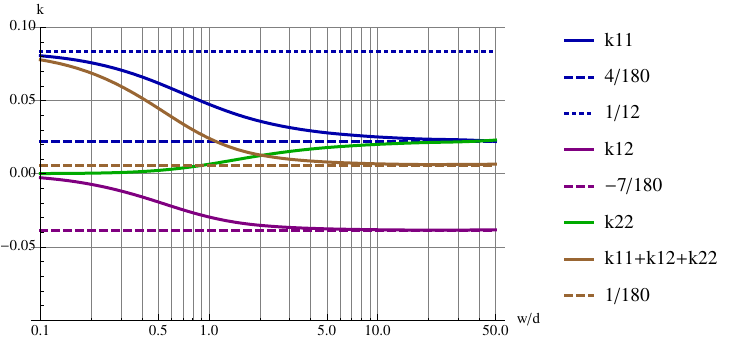}
  \caption{The coefficients $k_{11}, k_{12}, k_{22}$ for different values of $w/d$, where $w$ is the width of the square pad and $d$ is the silicon thickness. The dotted lines represent the for very small pads and very large pads as discussed in the text.}
 \label{k_coefficients}
  \end{center}
\end{figure}

\clearpage

\subsection{Leading edge discrimination \label{leading_edge_section}}

\begin{figure}
 \begin{center}
    a)
    \includegraphics[width=7.5cm]{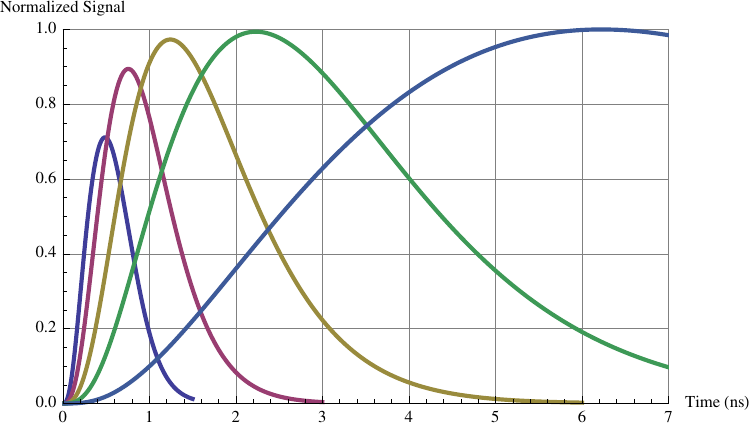}
    b)
    \includegraphics[width=7.5cm]{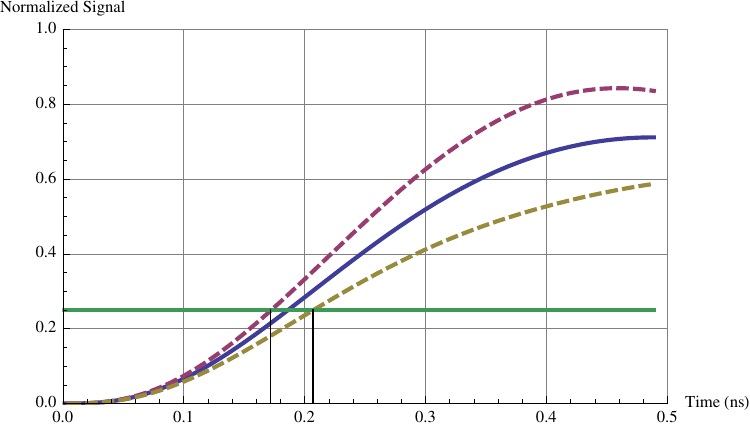}
    \caption{a) Average normalized signal $\ov h(t)$ for amplifier peaking times $t_p=0.25,0.5,1,2,6$\,ns for a 50$\mu$m sensor and V=$200\,$V. b) Average normalized signal $\ov h(t)$ for $t_p=0.25$\,ns together with the curves $\ov h(t)+\Delta_h(t)$ and $\ov h(t)-\Delta_h(t)$. }
  \label{threshold_discrimination_figure}
  \end{center}
\end{figure}

\begin{figure}
 \begin{center}
    a)
    \includegraphics[width=7.5cm]{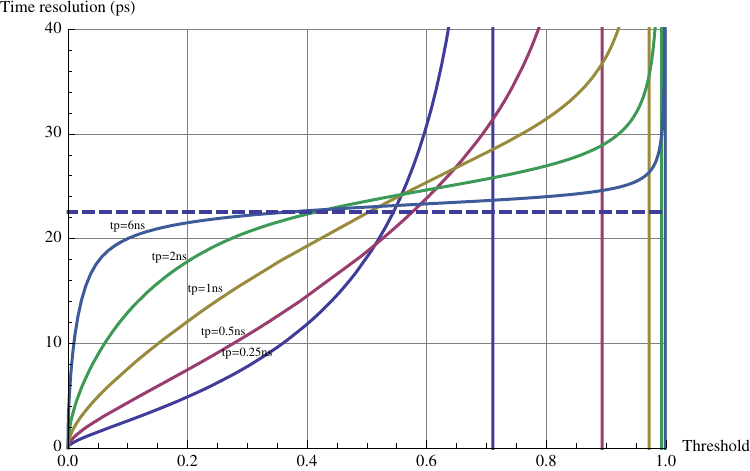}
    b)
    \includegraphics[width=7.5cm]{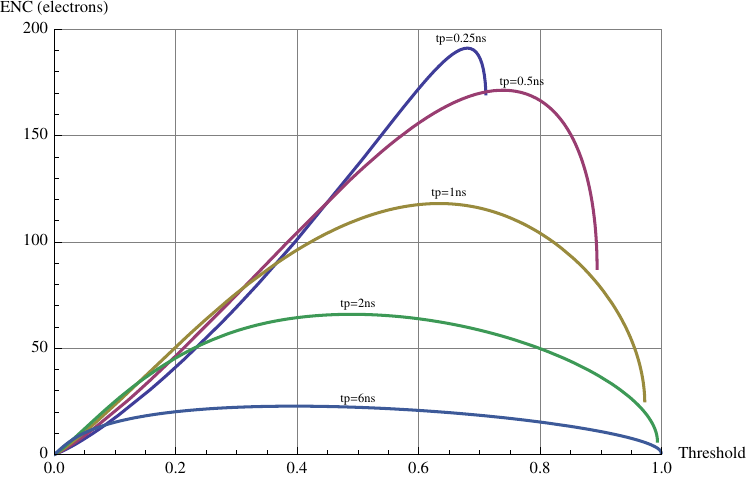}
  \caption{a) Time resolution for a sensor of 50$\mu$m thickness at 200\,V bias voltage. The slewing correction is performed by dividing the signal by the total charge and applying the threshold as a fraction of this charge. b) ENC needed to match the noise contribution to the effect from the Landau fluctuations.}
  \label{threshold_50um}
  \end{center}
\end{figure}

\begin{figure}
 \begin{center}
   a)
    \includegraphics[width=7.5cm]{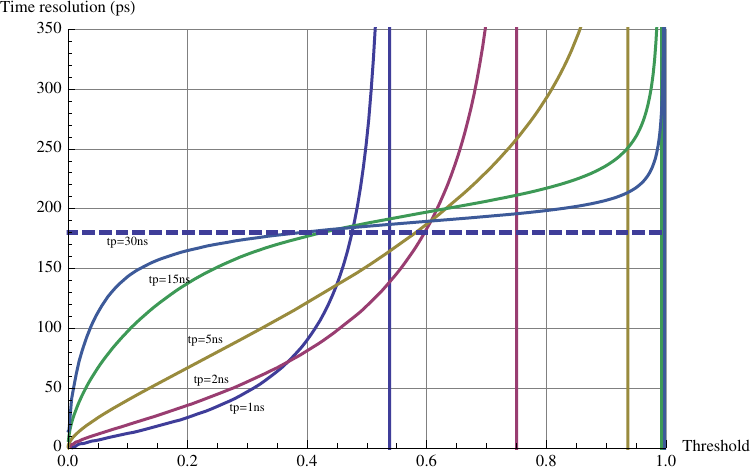}
   b)
    \includegraphics[width=7.5cm]{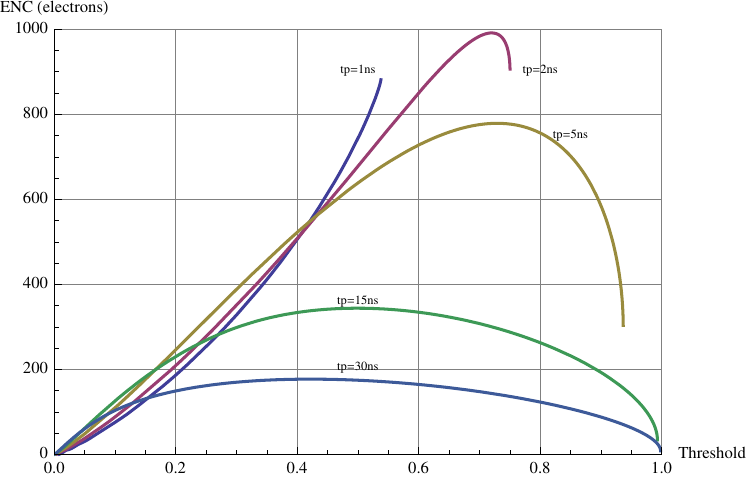}
  \caption{a) Time resolution for a sensor of 200$\mu$m thickness at 200\,V bias voltage. The slewing correction is performed by dividing the signal by the total charge and applying the threshold as a fraction of this charge. b) ENC needed to match the noise contribution to the effect from the Landau fluctuations.}
  \label{threshold_200um}
  \end{center}
\end{figure}

Up to this point we have just discussed the centroid time of the detector signals. In this section we consider the measured time to be determined by leading edge discrimination of the normalized detector signal. We process the detector signal by an amplifier of a given peaking time, and perform the so called 'slewing correction' for eliminating the timewalk effect from pulseheight fluctuations by dividing the amplifier output signal by the total signal charge and set the threshold to a given fraction of this signal. The current signal due to a single charge pair $-q, q$ at position $x,y,z$ is
\beq
   i_0(x, y, z, t) = -q \left[
   v_1 \,E_w(x,y,z-v_1t) \Theta(z/v_1-t) + v_2\,E_w(x,y,z+v_2t) \Theta((d-z)/v_2-t)
   \right]
\eeq
The current signal for having $n_1$ e/h pairs at $z=\Delta z$, $n_2$ e/h pairs at $z=2\Delta z$ etc. is given by
\beq
     i(n_1,n_2,...,n_N,x,y,t) = \sum_{k=1}^N n_k i_0(x,y,k\Delta z,t)
\eeq
We now process this signal by an amplifier with delta response $c f(t/t_p)$ where $t_p$ is the peaking time, $f(1)=1$, $c$ is the amplifier sensitivity in units of $[V/C]$ and $f(x)$ is defined by  
\beq
     f(x) = x^n\,e^{n(1-x)}
\eeq
The amplifier output signal is the given by the convolution of the induced signal and the amplifier delta response 
\bea \label{signal_notnormalized}
    s(n_1, n_2, ..., n_N, x,y,t)  & = & c\,\int_0^t f\left(  
        \frac{t-t'}{t_p}
      \right)
      \,i(n_1, n_2, ..., n_N, x,y,t') dt'  \\
      & = & c \,q
     \sum_{k=1}^N \,n_k \,g(x,y,k\Delta z,t)
\eea
where  $g(x,y,z,t)$ is 
\bea
  g(x,y,z,t) &=& \Theta ( z-v_1t) \int_{\frac{z-v_1t}{d}}^\frac{z}{d} 
  f\left( \frac{v_1t-z+ud}{v_1t_p}  \right) \, E_w^z(x/d,y/d,u,w_x/d,w_y/d,1)du  \\
   &+& \Theta ( v_1t-z) \int_{0}^\frac{z}{d} 
  f\left( \frac{v_1t-z+ud}{v_1t_p}  \right) \, E_w^z(x/d,y/d,u,w_x/d,w_y/d,1)du \no \\
     &+& \Theta [ (d-z)-v_2t] \int_{\frac{z}{d}}^{\frac{z+v_2t}{d}} 
  f\left( \frac{v_2t+z-ud}{v_2t_p}  \right) \, E_w^z(x/d,y/d,u,w_x/d,w_y/d,1)du \no \\
       &+& \Theta [ v_2t-(d-z)] \int_{\frac{z}{d}}^{1} 
  f\left( \frac{v_2t+z-ud}{v_2t_p}  \right) \, E_w^z(x/d,y/d,u,w_x/d,w_y/d,1)du  \no
\eea
The weighting field $E_w^z(x,y,z,w_x,w_y,d)$ for a pixel is given in Eq. \ref{weighting_field_final} of \ref{appendix_weightfield}. To perform slewing corrections we divide the signal by the total charge $q\sum n_k$ and we get the normalized amplifier output signal
\beq \label{normalized_signal}
     h(n_1, n_2, ..., n_N, x,y,t) = \frac{c}{\sum_{k=1}^N n_k } \,
     \sum_{k=1}^N \,n_k \,g(x,y,k\Delta z,t)
\eeq
The average normalized signal and the variance of the signal evaluate to 
\beq \label{average_leading_edge}
   \ov h(t) = \frac{c}{w_x\,w_y}\,\iint \left[\int_0^1  \, g(x,y,sd,t) ds\right]dx dy 
\eeq
and 
\bea  \label{variance_leading_edge}
  \Delta_h^2(t) &= &
 w(d)^2 \frac{c^2}{w_x w_y} 
    \iint \left[
      \int_0^1 g(x,y,sd,t)^2ds -
     \left(
     \int_0^1 g(x,y,sd,t)ds
     \right)^2
    \right]dx dy  \no  \\
    & + & 
    \frac{c^2}{w_x w_y} \iint  \left(  \int_0^1 g(x,y,sd,t) ds
     \right)^2 dx dy -
     \left[
     \frac{c}{w_xw_y} \iint  \left( \int_0^1 g(x,y,sd,t) ds \right)dx dy
     \right]^2  
\eea
The time resolution is then defined by (Fig. \ref{threshold_discrimination_figure}b)
\beq \label{leading_edge_discrimination_noise}
  \sigma_t = \frac{\Delta_h(t)}{\ov h'(t)}
\eeq
Here we just discuss the example of an infinitely extended pixel i.e. we use $E_w^z(x, y, z, w_x, w_y, d)=1/d$, which evaluates $g(x, y, z, t)$ to
\bea
    \frac{n^{n+1}}{e^n}\frac{d}{t_p}\,g(x,y,z,t) 
   & = & 
     v_1\, \Theta(z-v_1t) \, [n!-\Gamma(n+1,t/t_p)]  \no  \\ 
   &-& 
    v_1 \,\Theta(v_1t-z) \, [\Gamma(n+1,t/t_p)-\Gamma(n+1,-(z-v_1t)/(t_pv_1)] \no \\
   &+&
   v_2\, \Theta((d-z)-v_2t) \, [n!-\Gamma(n+1,t/t_p] \no \\
   &-& 
   v_2 \,\Theta(v_2t-(d-z)) \, [\Gamma(n+1,t/t_p)-\Gamma(n+1,-(d-z-v_2t)/(t_pv_2)]  \no    
\eea
where $n$ and $t_p$ are the parameters defining the amplifier. 
%If $n$ is an integer, the incomplete Gamma function $\Gamma(n+1,x)$ evaluates to
%
%
%
%\beq
%   \Gamma(n+1,x)= e^{-x}\,n! \sum_{k=0}^n \frac{x^k}{k!}
%\eeq 
%
%
%
As an example the average signal $\ov h(t)$  for a 50\,$\mu$m sensor at 200\,V for different peaking times is shown in Fig. \ref{threshold_discrimination_figure}a. The signal duration is around 0.8\,ns, so for small peaking times of 0.25 and 0.5\,ns there is significant 'ballistic deficit' while for peaking times $>1$\,ns the amplifier 'integrates' the full signal and the normalized amplitude becomes unity. In Fig. \ref{threshold_discrimination_figure}b the average normalized signal for a peaking time of 0.25\,ns is shown, together with $\pm 1$ standard deviations. \\
The resulting time resolution is shown in Fig. \ref{threshold_50um}a and Fig. \ref{threshold_200um}a for a 50\,$\mu$m and a 200\,$\mu$m sensor. We find that for large peaking times, the time resolution indeed approaches the centroid  time value, while for smaller peaking times the time resolution can be significantly better when setting the threshold at less than 30-40\% of the normalized signal. E.g. for the 50\,$\mu$m sensor at 200\,V, a peaking time of 0.25\,ns and a threshold set to 40\% of the total signal charge one should arrive at a resolution that is two times better than the resolution achieved with the centroid time. For a 200\,$\mu$m sensor, $t_p=5$\,ns and a threshold at 30\% of the signal one also expects a twice better resolution as compared to the centroid time.
\\ 
To study the impact of the noise we assume $\sigma_{noise}$ to be given in units of electrons. This noise is superimposed to the signal $s(t)$ from Eq. \ref{signal_notnormalized}, so when normalizing the signal to arrive at $h(t)$ we also have to normalize the noise by the total amount of charge deposited in the sensor. The average normalized noise the becomes 
\beq
     \overline{ \sigma}_{norm} = \int_0^\infty \frac{\sigma_{noise}}{n}p(n,d)\,dn = 
     \sigma_{noise}\,\frac{\lambda}{n_0\, d} \, \frac{1}{1+1.155\ln  d/\lambda}
\eeq
The contribution of the noise to the time resolution is then
\beq
    \sigma_t = \frac{\overline \sigma_{norm}}{\ov h'(t)}
\eeq
We can therefore express the required noise level when using a threshold of $\ov h(t)$, that matches the resolution from Landau fluctuations from Eq. \ref{leading_edge_discrimination_noise}, as
\beq
      \sigma_{noise}[electrons] = \Delta_h(t)\,\frac{n_0\,d}{\lambda}\,(1+1.155\ln d/\lambda)
\eeq
The numbers are shown in Fig. \ref{threshold_50um}b and Fig. \ref{threshold_200um}b. For the 50\,$\mu$m sensor and $t_p=0.25\,$ns the required noise level is 100 electrons and for the 200\,$\mu$m sensor at $t_p=5$\,ns the required noise is 400 electrons.

\clearpage

%%%%%%%%%%%%%%%%%%%%%%%%%%%%%%%%%%%%%%%%%%%%%%%%%%%%%
%%%%%%%%%%%%%%%%%%%%%%%%%%%%%%%%%%%%%%%%%%%%%%%%%%%%%

\section{Silicon sensors with internal gain}
\subsection{Centroid time resolution for silicon sensors with internal gain}

\begin{figure}
 \begin{center}
  \includegraphics[width=8cm]{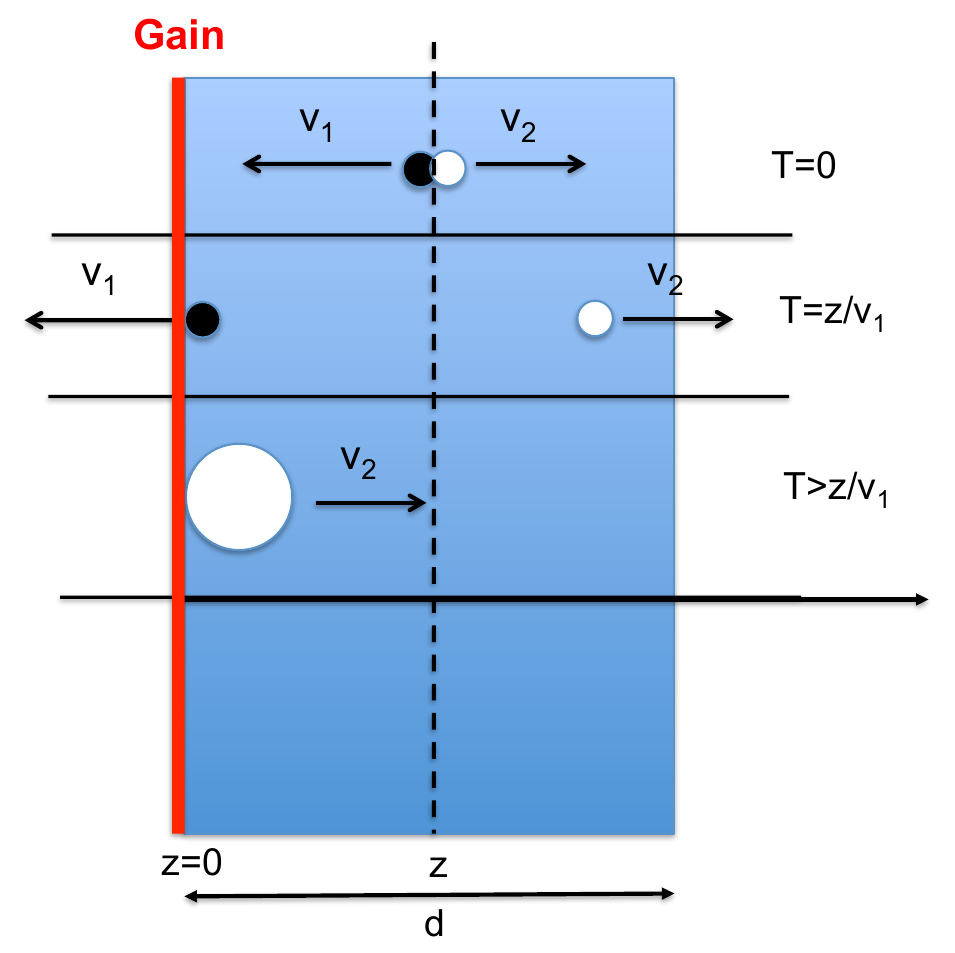}
  \caption{Silicon sensor with internal gain. An e-h par is produced at position $z$, the electron arrives at $z=0$ at time $T=z/v_1$, the electron multiplies in a high field layer at $z=0$ and the holes move back to $z=d$, inducing the dominant part of the current signal. }
  \label{sensor_slices_gain}
  \end{center}
\end{figure}
\begin{figure}
 \begin{center}
  \includegraphics[width=14cm]{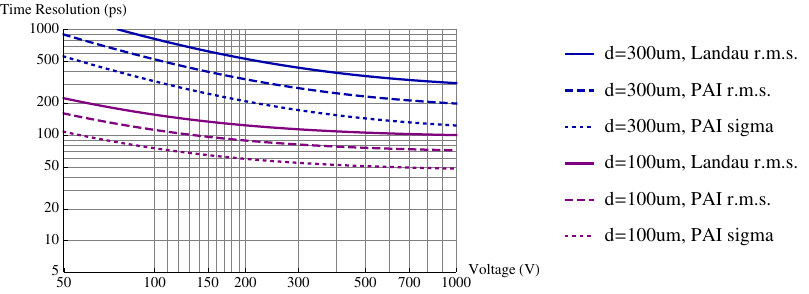}
    \includegraphics[width=14cm]{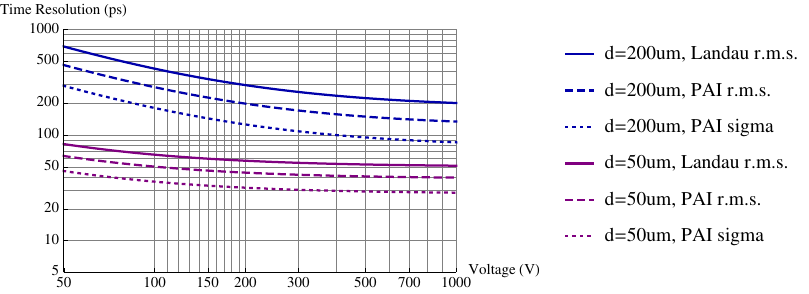}
  \caption{Time resolution for the centroid time from Eq. \ref{gain_resolution}  for 50, 100, 200, 300$\mu$m silicon  sensors  with internal gain of electrons, assuming a signal only from gain holes. The three curves for each sensor thickness correspond to the Landau theory, the PAI model and a Gaussian fit to the PAI model.}
  \label{reso_gain_figure}
  \end{center}
\end{figure}
In the Low Gain Avalanche Diode (LGAD), a high field region is implemented in the sensor in order to multiply electrons at some moderate gain and as a result improve the signal to noise ratio. We assume the geometry from Fig. \ref{sensor_slices_gain} with the amplification structure located at $z=0$. The electrons will therefore move from their point of creation to this structure, get multiplied and the holes created in the multiplication process are moving back from $z=0$ to $z=d$ through the entire sensor thickness $d$. If we assume 1) the gain $G$ to be sufficiently large such that the signal from the primary electron and hole movement is negligible, 2) the amplification structure to be infinitely thin, 3) a sensor with negligible depletion voltage, the signal from a single e-h pair created at position $z$ is of rectangular shape with duration $T=d/v_2$, shifted by the time $t=z/v_1$  
\beq
   i(t) = -G\,\frac{q\,v_2}{d}
   \left[
      \Theta(t-z/v_1)-\Theta(t-z/v_1-d/v_2)
   \right]
\eeq
The centroid  time of this signal is 
\beq
    \tau  = \frac{d}{2v_2}+\frac{z}{v_1}
\eeq
The centroid time for the case of $n_1, n_2, ..., n_N$ clusters at positions $z_1, z_2, ... , z_N$ is 
\beq
     \tau(n_1, n_2, ..., n_N)  = \frac{1}{\sum_{k=1}^N n_k}\, \sum_{k=1}^N \, 
     n_k 
     \left(
         \frac{d}{2v_2} +\frac{z_k}{v_1}  
     \right) =  \frac{d}{2v_2} +\frac{1}{\sum_{k=1}^N n_k}\, \sum_{k=1}^N \, 
     n_k   \frac{z_k}{v_1}
\eeq
The average and standard deviation of the centroid time are then 
\beq \label{gain_resolution}
   \ov \tau =\frac{d}{2}
   \left(
    \frac{1}{v_1}+\frac{1}{v_2}
   \right)
   \qquad
\Delta_{\tau} = w(d) \, \frac{d}{\sqrt{12}v_1} \approx
\frac{1}{\sqrt{a+b\ln d/\lambda+c (\ln d/\lambda)^2}}\,\frac{T_1}{\sqrt{12}}
\eeq
with $T_1=d/v_1$ being the total electron drift time. This expression is the same as the one from Eq. \ref{arrival_time_formula1} and  Eq. \ref{arrival_time_formula2}, so this sensor is simply measuring the arrival time distribution of the electrons at $z=0$. The resulting time resolution for 50, 100, 200, 300\,$\mu$m sensors is shown in Fig. \ref{reso_gain_figure}. Although the time resolution for the sensors with gain is worse than the one for silicon sensors without gain as shown in Fig. \ref{reso}, the big advantage of the sensors with gain is the improved signal to noise ratio that can 'eliminate' the effect from the noise. For a 50\,$\mu$m sensor at 220\,V one can achieve a time resolution of 30\,ps in accordance with measurements on the LGAD sensors.
 \\
The effects defining the time resolution for a sensor with gain therefore differ significantly from one without gain. The electrons first have to arrive at $z=0$ before being amplified and producing the gain signal, so the signal timing is defined by the arrival time distribution of the electron clusters at $z=0$. This is also illustrated by the fact that the second factor in Eq. \ref{gain_resolution} is simply the total transit time $T_e=d/v_1$ of the electrons through the full silicon thickness divided by $\sqrt{12}$. 

%%%%%%%%%%%%%%%%%%%%%%%%%%%%%%%%%%%%%%%%%%%%%
%%%%%%%%%%%%%%%%%%%%%%%%%%%%%%%%%%%%%%%%%%%%%
%%%%%%%%%%%%%%%%%%%%%%%%%%%%%%%%%%%%%%%%%%%%%

\subsection{Weighting field effect on the centroid  time for silicon sensors with gain}

\begin{figure}
 \begin{center}
  a)
  \includegraphics[height=3.2cm]{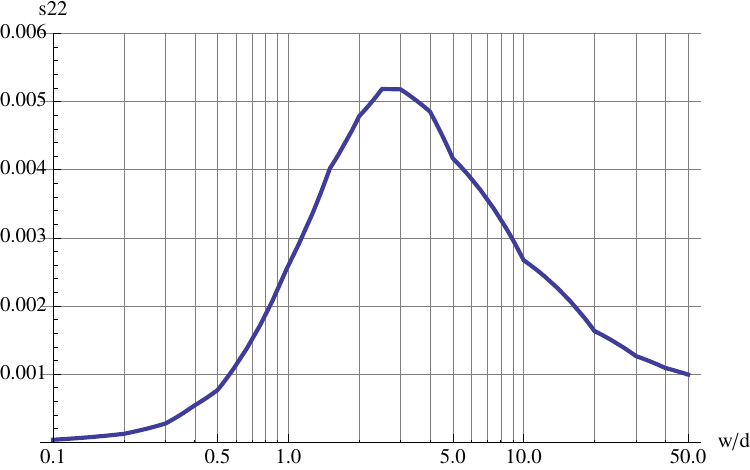}
  b)
  \includegraphics[height=3.2cm]{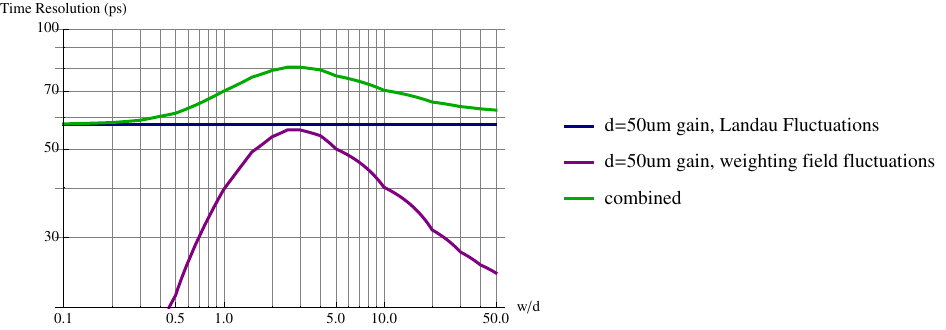}
  \caption{a) Coefficient $s_{22}$ defining the impact of the weighting field on the time resolution. b) Centroid  time resolution for a gain sensor of 50\,$\mu$m thickness at 200\,V. The horizontal line shows the contribution from Landau fluctuations only, while the other lines show the contribution from weighting field fluctuations as well as the combined effect.  }
  \label{reso_gain_weightfield}
  \end{center}
\end{figure}

In this section we discuss the effect of the finite pixel size on the centroid  time resolution for sensors with gain. Assuming the readout electrode at $z=0$ to be segmented into pixels with an associated weighting potential $\phi_w(x,y,z)$, the induced signal due to a single charge pair created at position $x, y, z$ at $t=0$ becomes 
\beq
   i(t) = -G\,q\,v_2\,E_w[x, y, v_2(t-z/v_1)]
   \left[
      \Theta(t-z/v_1)-\Theta(t-z/v_1-d/v_2)
   \right]
\eeq
and the centroid  time for this signal is given by
\beq \label{tauxyz_gain}
  \tau(x, y, z) =  \frac{z}{v_1}+\frac{d}{v_2} \int_0^1 \phi_w(x, y, s\,d)ds
\eeq
Assuming a uniform charge deposit along the track, the centroid  time becomes
\beq
   \tau(x, y) = \frac{1}{d}\int_0^d \tau(x, y, z) dz = \frac{d}{2v_1} +\frac{d}{v_2} \int_0^1 \phi_w(x, y, s\,d)ds
\eeq
The variance for uniform irradiation of the pad is then
\bea \label{weighting_field_resolutions}
    \Delta_\tau^2 & = & \ov{ \tau^2}-\ov \tau ^2  \no \\
    &= & \frac{d^2}{v_2^2}\left[
   \frac{1}{w_x w_y} \iint 
   \left(
    \int_0^1 \phi_w(x, y, s\,d)ds
   \right)^2 dx dy-
   \left(
    \frac{1}{w_x w_y}\iint 
   \left(
    \int_0^1 \phi_w(x, y, s\,d)ds
   \right)dx dy
   \right)^2
   \right] \no \\
   &=&  \frac{d^2}{v_2^2}\,s_{22} = T_2^2 \, s_{22}
\eea
which is the pendant to Eq. \ref{weighting_field_resolution1} for sensors without gain. The coefficient $s_{22}$ for different pixel sizes is are listed in Teable \ref{sstable} and shown in Fig. \ref{reso_gain_weightfield}a. The effect on the time resolution for a $50\,\mu$m sensor is shown in Fig. \ref{reso_gain_weightfield}b. The effect is again largest for pixel sizes of $w/d \approx 3$. In case we also take into account the Landau fluctuations we have to use Eq. \ref{tauxyz_gain} in Eq. \ref{variance_combined} and find 
\beq
    \Delta_\tau^2 =\ov{ \tau^2}-\ov \tau ^2 = 
   w(d)^2\frac{d^2}{12\,v_1^2} + \frac{d^2}{v_2^2}s_{22} =  
    w(d)^2\,\frac{T_1^2}{12}+T_2^2s_{22}
\eeq
which is the pendant to Eq. \ref{weighting_field_resolutionk} for sensors without gain. So we find the interesting result that for this case there is no correlation between the Landau fluctuations and the weighting field fluctuations, and the two components just add in squares. We also note that the result will be the same whether we segment the electrode at $z=0$ where the multiplication takes place or whether we segment the electrode at $z=d$.

\subsection{Impact of gain fluctuations}

The electron amplification in the gain layer of the LGAD will have statistical fluctuations and in the following we want to quantify the impact of these fluctuations. In case the amplification process is such that the ionizing collisions are independent and do not have a history to the previous collision, the fluctuations of the gain for a single electron are governed by the Yule-Furry law according to 
\beq
       p(G)=\frac{1}{\ov G} \left(1-\frac{1}{\ov G}\right)^{G-1} \qquad \Delta_G^2=\ov G(\ov G-1)\approx \ov G^2
\eeq 
where $\ov G$  is the average gain. This assumption is correct as long as the fields are sufficiently low such that there is only electron multiplication and the multiplication of holes is negligible.
In case there are $n \gg 1$ primary electrons, the distribution of the number of electrons after multiplication will assume a Gaussian shape with $\mu = n G$ and $\sigma^2 = n \Delta_G^2=n\ov G^2$ due to the central limit theorem. The resulting charge spectrum is therefore a convolution of this Gaussian with the Landau distribution $p(n,d)$. To estimate the effect of the gain fluctuations on the Landau distribution we approximate the Landau distribution with a Gaussian of mean and standard deviation according to
\beq
      \mu = n_{MP} \qquad \sigma=\frac{\Delta n_{FWHM}}{2\sqrt{2 \ln 2}}
\eeq
The convolution of this Gaussian with the Gaussian from the gain fluctuations will then again result in a Gaussian where the variances are added in squares and we have
\beq
    \frac{\Delta n_{FWHM}^G}{n_{MP}} =   \frac{\Delta n_{FWHM}}{n_{MP}}
    \sqrt{1+\frac{n_{MP}8\ln 2 }{\Delta n_{FWHM}^2}} 
    \approx
    \frac{\Delta n_{FWHM}}{n_{MP}}
   \left(1+\frac{n_{MP}4\ln 2 }{\Delta n_{FWHM}^2}  \right)
   =
      \frac{\Delta n_{FWHM}}{n_{MP}} \left(1+\varepsilon  \right)
\eeq
The value of $\varepsilon$ ranges from $1.9\times 10^{-3}$ for $d=50\,\mu$m to 
$4.1\times 10^{-4}$ for $d=300\,\mu$m. The gain fluctuations will therefore increase the relative fluctuations of the charge deposit by less than 0.2\% for a 50\,$\mu$m sensor and even less for the 300\,$\mu$m sensor. \\
The correct resulting charge distribution $p_G(n,d)$ when assuming the Landau distribution $p(n,d)$ for the primary charge deposit is given by
\beq
   p_G(n,d) = \frac{1}{G} \int_0^\infty p(m,d)\frac{1}{\sqrt{2 \pi m} }
   \exp \left(
   -\frac{(n/G-m)^2}{2m}
   \right)dm
\eeq
and the evaluation is shown in \ref{appendix_gain1}. The correct values of $\epsilon$ for the increase of the FWHM with respect to the original distribution are $2.8/1.6/0.86/0.61\times 10^{-3}$  for the 50/100/200/300\,$\mu$m sensor. \\
In order to evaluate the impact on the time resolution we have to find the effective cluster size distribution $p_{clu}^G(n)$. For large numbers of $\ov G$, the Furry law turns into the exponential distribution 
\beq
   p(G) = \frac{1}{\ov G} \,e^{-G/\ov G}
\eeq
Even for the typically low LGAD gains of about 20 this is a good approximation. The probability to find $n$ electrons for $m$ primary electrons is then given by the $n$-times self convolution of this expression and we have
\beq
    p(n) = \frac{1}{\ov G} \,e^{-n/\ov G}\,\frac{1}{(m-1)!}\left( \frac{n}{\ov G}\right)^{m-1}
\eeq
The effective cluster size distribution for $p_{clu}(n)=n_0/n^2\,\Theta(n-n_0)$ is then 
\beq
    p_{clu}^G(n) = \frac{1}{\ov G} \,e^{-n/\ov G} \int_{n_0}^\infty 
    \frac{n_0}{m^2 \Gamma(m)} \left( \frac{n}{\ov G}\right)^{m-1} dm
\eeq
Using this effective cluster size distribution together with the distribution $p_G(n,d)$ in Eq. \ref{general_xd} we can evaluate the impact on the time resolution and have
\beq
     \frac{\Delta_\tau^G}{\Delta_\tau} = \sqrt{\frac{w_G(d)}{w(d)}}=1+\varepsilon
\eeq
where $\varepsilon = 9/4.6/2.2/1.5\times10^{-4}$. The effect of gain fluctuations on the time resolution is less than 0.1\,\% for sensors of more than $50\,\mu$m thickness and is therefore completely negligible.

%%%%%%%%%%%%%%%%%%%%%%%%%%%%%%%%%%%%%%%%%%

\subsection{Leading edge discrimination for silicon sensors  with gain}

\begin{figure}
 \begin{center}
 a)
    \includegraphics[width=7.5cm]{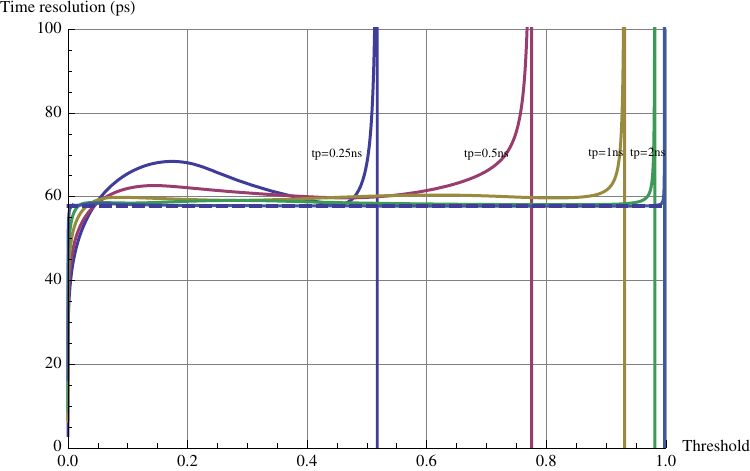}
    b)
    \includegraphics[width=7.5cm]{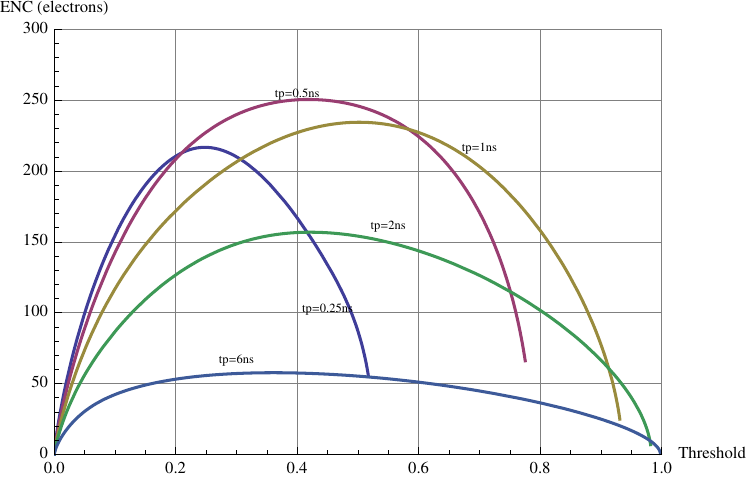}
  \caption{a) Time resolution for a gain sensor of 50$\mu$m thickness at 200\,V bias voltage when applying a threshold to the signal normalized by the total charge, assuming the Landau theory. The values do not improve beyond the centroid  time resolution which is indicated by the dashed horizontal line.
  b) ENC needed to match the noise effect of the time resolution to the effect from the Landau fluctuations.}
  \label{gain_leading_edge_plot}
  \end{center}
\end{figure}

In this section we discuss the time resolution when considering leading edge discrimination of sensors with gain. We proceed as in Section \ref{leading_edge_section}  and convolute the signal from a single e-h pair at position $z$ 
\beq
   i_0(x, y, z, t) = -G q v_2 E_w(x, y, v_2(t-z/v_1))
   \left[
      \Theta(t-z/v_1)-\Theta(t-z/v_1-d/v_2)
   \right]
\eeq
with the electronics delta response and find
\bea
  g(x, y, z, t) & = & \Theta(t-z/v_1)\Theta(d/v_2+z/v_1-t)
  \int_0^{\frac{v_2}{d}(1-\frac{z}{v_1})}
  f\left(\frac{t-z/v_1-ud/v_2}{t_p} \right)
   E\left( \frac{x}{d},\frac{y}{d},u,\frac{w_x}{d},\frac{w_y}{d},1\right) du \no \\
  & + &
\Theta(t-d/v_2-z/v_1)
  \int_0^{1}
  f\left(\frac{t-z/v_1-ud/v_2}{t_p} \right)
   E\left( \frac{x}{d},\frac{y}{d},u,\frac{w_x}{d},\frac{w_y}{d},1\right) du 
\eea
which for an infinitely extended electrode with $E_w=1/d$ evaluates to
\bea
    \frac{n^{n+1}}{e^n}\frac{d}{t_p}\,g(x,y,z,t) & = &v_2
     \Theta(t-z/v_1)\Theta(d/v_2+z/v_1-t)
     \left[
       n! -  \Gamma\left( 
       n+1,\frac{n(v_1t-z)}{t_pv_1}
       \right)
     \right]  \\
     &-& v_2
    \Theta(t-d/v_2-z/v_1)
    \left[
        \Gamma \left(
        n+1,\frac{n(v_1t-z)}{t_pv_1}
        \right)
        - \Gamma \left(
        n+1,\frac{n(t-d/v_2-z/v_1)}{tp}
        \right)
    \right] \no
\eea
Evaluating Eq. \ref{average_leading_edge}, Eq. \ref{variance_leading_edge} and Eq. \ref{leading_edge_discrimination_noise} we then find the results shown in Fig. \ref{gain_leading_edge_plot}a. We find that even for leading edge discrimination of the normalized signal the time resolution for a sensor with gain does not improve beyond the centroid  time resolution value. The reason is that in the outlined formulas the signal is normalized by the total charge deposited in the sensor. The signal that makes up the leading edge has however no correlation with the total deposited charge but is only related to the number of electrons that have already arrived at the gain layer. This is very different from the standard silicon sensor without gain, where the movement of all deposited charges  makes up the leading edge signal. \\
If one want wants to improve the time resolution of silicon sensors with gain beyond the centroid time resolution, one therefore needs ultra fast front-end electronics with slewing corrections related to the leading edge of the signal and not to the total charge of the signal. This goes beyond the mathematical formalisms developed in this report and Monte Carlo simulations have to be used to study this scenario.

\medskip

\section{Comparison with measurements}

In \cite{Cartiglia201783} the time resolution of an LGAD sensor with 50\,$\mu$m thickness is quoted as $\sigma = 34$\,ps at 200\,V and $\sigma = 27$\,ps at 230\,V. Eq. \ref{gain_resolution}  predicts a centroid time resolution of $\sigma = 32$\,ps for 200\,V and $\sigma = 31$\,ps for 230\,V for the PAI model. The measured and calculated numbers are therefore in the same range, which seems to confirm the effect shown in Fig. \ref{gain_leading_edge_plot}, namely that even when using leading edge discrimination with electronics of $\approx 0.5$\,ns peaking time for this sensor one is effectively measuring the centroid time. \\ \\
In \cite{Akchurin2017} the time resolution for multiple particles passing a 133, 211, 285\,$\mu$m sensor is given. All sensors were biased at 600\,V. An amplifier delta response of $1$\,ns peaking time is used, resulting in a peaking time for the average signal of the 211\,$\mu$m sensor of $\approx 2$\,ns. Leading edge discrimination at 50\% of the signal peak is used. Eq. \ref{final_formula} predicts centroid time resolutions of $\sigma = 24, 41, 60$\,ps for the three sensors when using the PAI model. With a peaking time of 1\,ns and the threshold set at 50\,\% of the signal Eq. \ref{leading_edge_discrimination_noise} predicts a resolution of $\sigma \approx 14$\,ps, for all three values of sensor thickness. From Eq. \ref{multiple_particle_formula} we see that the scaling factor when having 100 MIPs instead of one MIP amounts to $\approx 0.77$, so we expect a time resolution of 11\,ps for all these cases, which actually does approximately match the quoted number where the resolution saturates.
\\ \\
The NA62 Gigatracker uses a 200$\,\mu$m sensor with $300 \times 300$\,$\mu$m pixels. The signals are read by a frontend with 5\,ns peaking time and the threshold is set to around 30\% of the signal. A measured time resolution of 190\,ps for 200\,V is quoted \cite{AglieriRinella2017147}. The effect of noise on these numbers is quoted to be negligible. To compare to calculations, we would  in principle have to evaluate Eq. \ref{variance_leading_edge}  for leading edge discrimination of a sensor with finite pixel size, which turns out to be unfeasible, so we compare to some limiting cases. The PAI model and leading edge discrimination at about 35\,\% of the signal for 200\,V predicts a time resolution of 64\,ps r.m.s. (42\,ps $\sigma$) for an infinitely large pad. The observed time resolution is therefore dominated by the weighting field effect. 
The impact of the centroid time for the weighting field (correlated with the Landau fluctuations) effect is 272\,ps r.m.s. (224\,ps $\sigma$). The effect of leading edge discrimination on the weighting field effect, which is not discussed in this report, will reduce this number to some extent, so the measured 190\,ps are in the right ballpark.  For a more accurate quantitative evaluation, a Monte Calo simulation must be performed.
\\ \\
In \cite{100ps_timing} a time resolution of 100\,ps is reported for a sensor of 100\,$\mu$m thickness and $800\times800\,\mu$m pixels, biased at 230\,V. An amplifier of 200-400\,ps rise-time is used and a time resolution of 100\,ps is reported. The PAI model predicts a centroid time resolution of $\sigma = 26$\,ps for this sensor, and the leading edge discrimination  will still result in some improvement on top of this number. As shown in the paper, the time resolution is fully dominated by the noise contribution, so we cannot extract the time resolution component due to Landau fluctuations from this measurement.

%\clearpage

%%%%%%%%%%%%%%%%%%%%%%%%%%%%%%%%%%%%%%%%%%%%%
%%%%%%%%%%%%%%%%%%%%%%%%%%%%%%%%%%%%%%%%%%%%%
%%%%%%%%%%%%%%%%%%%%%%%%%%%%%%%%%%%%%%%%%%%%%

\section{Conclusions}

%We derived analytic expressions for the time resolution of silicon sensors.
\begin{itemize}

\item The probability for a relativistic particle to deposit $n$ e-h pairs in a silicon sensor of thickness $d$ is given by
\beq 
   p(n, d) = {\cal L}^{-1}
   \left[
   e^{d/\lambda(P_{clu}(s)-1)} 
   \right]
\eeq
where $P_{clu}(s)$ is the Laplace transform of the cluster size distribution and $\lambda$ is the average distance between primary collisions, which evaluates to $\lambda \approx$\,0.212\,$\mu$m for relativistic particles in silicon. For a $1/n^2$ cluster size distribution this expression becomes the Landau distribution, while for a more realistic cluster size distribution from the PAI model we get a distribution with a relative width that is 25-35\% smaller than the one from the Landau distribution.

\item The standard deviation of the centroid time of a silicon detector signal is given by 
\beq
   \Delta_\tau = w(d/\lambda)
   \sqrt{\frac{4}{180}T_{1}^2-\frac{7}{180}T_{12}^2+\frac{4}{180}T_{2}^2}
\eeq
assuming a large readout electrode and negligible depletion voltage. $T_1=d/v_1, T_2=d/v_2, T_{12}=d/\sqrt{v_1v_2}$ are the drift times of the electrons and holes. 
Using the Landau theory for charge deposit, the expression $w(d/\lambda)$ approaches $1/\sqrt{\ln d/\lambda}$ for large values of $d$. In the interval of $25<d<500\,\mu$m, $w(d/\lambda)$ can be approximated by
\beq
   w(d/\lambda) \approx \frac{1}{\sqrt{a+b\,\ln d/\lambda+c\,(\ln d/\lambda)^2}}
\eeq
with $a= 1, b=1.155, c=0$ for the Landau theory, $a=13.7, b=-4.9, c=0.85$ for a PAI charge deposit model and $a=47.7, b=-22.8, c=3.37$ when performing a Gaussian fit to the measured time distribution for the PAI model. \\
For a silicon sensor of 300\,$\mu$m thickness and 600\,V this evaluates to a resolution of 161, 103, 64\,ps, indicating that the Landau theory overestimates the fluctuations and that we have to clearly distinguish the r.m.s. and the Gaussian fit due to significant tails in the distribution.  For a 200\,$\mu$m sensor at 300\,V the resolution evaluates to 132, 88, 56\,ps. For a 50\,$\mu$m sensor at 200\,V the values are 22, 17, 12\,ps.

\item For multiple particles passing the silicon sensor the time resolution scales from the single particle time resolution $\Delta_{\tau}$(1 particle) as 
\beq
   \frac{\Delta_{\tau}(n\, \mbox{particles})}{\Delta_{\tau}(1\, \mbox{particle})} = 
   \frac{1}{\sqrt{1+\frac{\ln n}{\ln d/\lambda}}} 
\eeq 
which amounts to an improvement of only 26, 24, 23, 22\% for a 50, 100, 200, 300\,$\mu$m sensor when going from 1 to 100 particles.

\item Measuring the sensor signal with an amplifier of peaking time $t_p$ larger than the drift time of electrons and holes, the amplifier output is equal to the delta response, scaled by the total signal charge and shifted by the centroid  time. To determine the time of this pulse of known shape one can then use standard techniques of constant fraction discrimination and optimum filtering to extract the time information. Assuming the Landau theory, the average contribution of the noise to the time resolution is then
\beq
  \ov \sigma_t = \sigma_{noise}[electrons]\frac{\lambda}{d\,n_0}\,\frac{1}{1+1.155\ln d/\lambda}\,t_p\,c(n_s)
\eeq
where $t_p$ is the peaking time of the amplifier and $c(n_s)$ is a constant depending on the measurement technique. Using constant fraction discrimination at the maximum slope of the signal we have $c(n_s) \approx 0.55-0.6$. Using continuous signal sampling and optimum filtering one arrives at similar numbers when sampling at an interval of $t_p/2$ and one can achieve $c(n_s) \approx  0.2-0.3$ for very high frequency sampling. For $t_p=2$\,ns, $d=50\,\mu$m and an Equivalent Noise Charge (ENC) of 50 electrons we have a contribution from the noise of $\sigma_t \approx 17$\,ps, that has to be added in square with the numbers from Landau fluctuations. In order to exploit the intrinsic time resolution of thin silicon sensors one therefore needs ultra low noise performance of the frontend electronics. For a given series noise voltage $e_n$ of an amplifier, the equivalent noise charge decreases with $1/\sqrt{t_p}$, the effect of the noise on time resolution does however increase linearly with $t_p$. It is therefore advantageous to use faster electronics if power consumption allows and other noise sources do not start to become dominant.

\item Assuming a square readout pixel of dimension $w$, the variation of the track position and therefore the variation of the weighting field and related signal shape will have an impact on the time resolution and the standard deviation of the centroid  time becomes
\beq
    \Delta_\tau = 
    \sqrt{
   w(d/\lambda)^2(k_{11}T_1^2+k_{12}T_{12}^2+k_{22}T_2^2)
    +\left( c_{11}T_1^2+c_{12}T_{12}^2+c_{22}T_2^2 \right)
    }
\eeq 
Neglecting charge fluctuations and assuming a uniform charge deposit, the coefficients $k_{11}, k_{12}, k_{22}$ vanish. Assuming very large readout pixels, the coefficients $c_{11}, c_{12}, c_{22}$ vanish and $k_{11}, k_{12}, k_{22}$ become $4/180, -7/180, 4/180$ in accordance with the above. For very small pixels, we have $k_{11}=1/12$ and all other coefficients vanish, which is in accordance with an arrival time distribution of charges at the pad. Landau fluctuations and weighting field fluctuations are strongly correlated, so they cannot be decoupled or 'added in squares'. Since $k_{11}>k_{22}$, the effect of weighting field fluctuations is smallest if $T_1$ is small i.e. if the electrons move towards the readout pixel. In this case it seems possible that for values of $w/d \gtrsim 1$ the weighting field effect does not add significantly to the centroid  time resolution. We note that this calculation assumes perpendicular tracks and neglects diffusion.

\item The expressions for leading edge discrimination of the normalized silicon sensor signal (i.e. the signal divided by the total charge) show that the centroid  time resolution is indeed recovered for large peaking times, and that for faster electronics the time resolution is significantly improved when placing the threshold at $<40\%$ of the total signal charge. As an example, for a 50\,$\mu$m sensor at 200\,V, a peaking time of 1\,ns and a threshold at 30\,\% of the normalized signal, the time resolution improves by a factor 2 with respect to the centroid time and the noise must be less than $70$ electrons in order to not significantly add to this value. 
%\item The effect of the finite pixel size on the leading edge discrimination of the normalized signal can be calculated with the formulas given in this report, the numerical evaluation is however quite involved and a Monte Carlo simulation might be more efficient.

\item For silicon sensors with internal gain (LGAD), the standard deviation of the centroid  time becomes
\beq
   \Delta_\tau = w(d/\lambda) \,  \frac{T_1}{\sqrt{12}}
\eeq
This formula assumes that only the gain holes contribute to the signal. This expression is the same as the one for the very small pixels without gain and represents in essence an arrival time distribution. For a 200$\mu$m sensor at 300\,V the time resolution is 255, 170, 108\,ps for the Landau, PAI and Gaussfit PAI model. These numbers are a factor 2 larger compared to the sensor without gain. For a $50\,\mu$m sensor at 200\,V the numbers are 57, 44, 32\,ps, about a factor 2.5 larger than for the sensor without gain. The very big advantage of sensors with gain is the large signal to noise ratio that can make the noise contribution to the time resolution negligible and therefore allows large pixels, electronics with modest noise performance and modest bandwidth.

\item The impact of gain fluctuations on the time resolution for sensors with internal gain (LGAD) of 50-300\,$\mu$m thickness is on the 0.1\,\% level and therefore negligible.

\item Including the effect of the finite pixel size on the centroid  time resolution of a silicon sensor with gain we find
\beq
    \Delta_{\tau}=\sqrt{w(d/\lambda)^2 \,  \frac{T_1^2}{12} + s_{22}T_2^2}
\eeq 
In contrast to sensors without gain there is no correlation between the Landau fluctuations and the weighting field fluctuations.  For uniform charge deposit, only the second term of the expression remains. For very large and very small pads the coefficient $s_{22}$ vanishes and the effect is largest for $w/d \approx 3$. In addition the expression is the same, whether the electrode at the side of the gain layer $z=0$ or the electrode on the opposite side is segmented into pixels. 

\end{itemize}
The calculations presented in this report provide insight into some principle dependencies for the time resolution of silicon sensors on charge fluctuations, noise and weighting field fluctuations. The inclusion of more detailed models including the effect of diffusion, track angle, finite depletion voltage and pixelization are best accomplished through Monte Carlo simulations and the formulas of this report can be used as benchmarks for such studies.

\section*{Acknowledgement}

We would like to thank Heinrich Schindler for providing the data of the PAI model as well as Nicolo Cartiglia, Matthew Noy and Angelo Rivetti for important discussions.

%%%%%%%%%%%%%%%%%%%%%%%%%%%%%%%%%%%%%%%%%%%%%
%%%%%%%%%%%%%%%%%%%%%%%%%%%%%%%%%%%%%%%%%%%%%
%%%%%%%%%%%%%%%%%%%%%%%%%%%%%%%%%%%%%%%%%%%%%

\appendix

\section{\label{appendix_landau} }

Evaluating Eq. \ref{general_pn_distribution} with the specific model of the $1/n^2$ distribution from Eq. \ref{Landau_approximation} we find the Landau distribution $L(x)$ according to 

\bea
   L(x) & = & \frac{1}{2\pi i} \int_{\sigma - i\infty}^{\sigma+i\infty}
     \exp
     \left[   
           s x+s\ln s
      \right]
      ds \\
            & = & 
               \frac{1}{\pi} \int_0^\infty      
               \exp (-\pi/2\,t)
  \cos (tx+t\ln t)  dt \label{Landau_low} \\
   &= & 
    \frac{1}{\pi} \int_0^\infty    \exp
     \left[   
           -t x-t\ln t
      \right] \sin (\pi t)
      dt \label{Landau_high} 
\eea
Expression \ref{Landau_low}  is well suited for evaluation for $x<0$, while Eq. \ref{Landau_high} is well suited for evaluation for $x>0$. For large values of $x$ the Landau distribution approximates to 
\beq
   L(x) \approx \frac{1}{x^2}
\eeq

%%%%%%%%%%%%%%%%%%%%%%%%%%%%%%
%%%%%%%%%%%%%%%%%%%%%%%%%%%%%

\section{\label{appendix_statistics} }

The centroid time of the silicon detector signal assuming $n_k$ e-h pairs in slice $k$ is 
\beq 
   \tau(n_1, n_2, ..., n_N) = 
   \frac{1}{2d\,(\sum_{k=1}^N n_k)}\,
   \sum_{k=1}^N n_k
   \left[
   \frac{z_k^2}{v_1}+\frac{(d-z_k)^2}{v_2}
   \right]
\eeq
The average cetroid time $\ov \tau$ is then given by
\beq
    \ov \tau = \int_0^{\infty}\int_0^{\infty} ... \int_0^{\infty}  \tau(n_1, n_2, ..., n_N) 
    p(n_1,\Delta z)p(n_2, \Delta z)...p(n_N,\Delta z)\, dn_1\,dn_2...dn_N
\eeq
Since 
\beq
    \int_0^{\infty}\int_0^{\infty} ... \int_0^{\infty}  \frac{n_1+n_2+...+n_N}{n_1+n_2+...+n_N} 
    p(n_1,\Delta z)p(n_2, \Delta z)...p(n_N,\Delta z)\, dn_1\,dn_2...dn_N = 1
\eeq
we have
\beq
    \int_0^{\infty}\int_0^{\infty} ... \int_0^{\infty}  \frac{n_k}{n_1+n_2+...+n_N} 
    p(n_1,\Delta z)p(n_2, \Delta z)...p(n_N,\Delta z)\, dn_1\,dn_2...dn_N = \frac{1}{N} \qquad k=1, 2,..., N
\eeq
and therefore
\beq \label{tau_average_parallel_plate}
    \ov \tau= 
   \frac{1}{2d}
   \sum_{k=1}^N \frac{1}{N}
   \left[
   \frac{z_k^2}{v_1}+\frac{(d-z_k)^2}{v_2}
   \right]
   \approx
   \frac{1}{2d^2}\, \int_0^d    
   \left[
   \frac{z^2}{v_1}+\frac{(d-z)^2}{v_2}
   \right] dz = \frac{d}{6} 
   \left( 
   \frac{1}{v_1}+\frac{1}{v_2}
   \right)
\eeq
which is the expected centroid time of the two triangular signals form the electrons and the holes. The second moment of the centroid  time  $\ov {\tau^2}$ is given by
\beq
    \ov {\tau^2} = \int_0^{\infty}\int_0^{\infty} ... \int_0^{\infty}  \tau^2 (n_1, n_2, ..., n_N) 
    p(n_1,\Delta z)p(n_2, \Delta z)...p(n_N,\Delta z)\, dn_1\,dn_2...dn_N
\eeq
\beq
   \tau^2(n_1, n_2, ... , n_N) = 
   \frac{1}{4d^2\,(\sum_{k=1}^Nn_k)^2} \sum_{k=1}^N\sum_{r=1}^N n_k n_r
   \left[
   \frac{z_k^2}{v_1}+\frac{(d-z_k)^2}{v_2}
   \right]
   \left[
   \frac{z_r^2}{v_1}+\frac{(d-z_r)^2}{v_2}
   \right]
\eeq
We define
\bea
    a_N &=& \int_0^{\infty}\int_0^{\infty} ... \int_0^{\infty}  \frac{n_k\,n_r}{(n_1+n_2+...+n_N)^2}  
    p(n_1,\Delta z)p(n_2, \Delta z)...p(n_N,\Delta z)\, dn_1\,dn_2...dn_N \qquad k \neq r  \no \\
   b_N&=&\int_0^{\infty}\int_0^{\infty} ... \int_0^{\infty}  \frac{n_k^2}{(n_1+n_2+...+n_N)^2}  
    p(n_1,\Delta z)p(n_2, \Delta z)...p(n_N,\Delta z)\, dn_1\,dn_2...dn_N 
\eea
and since we have
\beq
    \int_0^{\infty}\int_0^{\infty} ... \int_0^{\infty}  \frac{(n_1+n_2+...+n_N)^2}{(n_1+n_2+...+n_N)^2} 
    p(n_1,\Delta z)p(n_2, \Delta z)...p(n_N,\Delta z)\, dn_1\,dn_2...dn_N = 1
\eeq
it holds that
\beq
     N\,b_N+N(N-1)a_N=1 \quad \rightarrow \quad 
     a_N = \frac{1-N\,b_N}{N(N-1)} \approx \frac{1}{N^2}-\frac{b_N}{N}
\eeq
The second moment of $\tau$ therefore becomes
\bea
   \ov{ \tau^2 } & = & 
   \frac{b_N}{4d^2} \sum_{k=1}^N 
   \left[
   \frac{z_k^2}{v_1}+\frac{(d-z_k)^2}{v_2}
   \right]^2+
    \frac{a_N}{4d^2}
    \sum_{k=1}^N\sum_{r\neq k=1}^N 
   \left[
   \frac{z_k^2}{v_1}+\frac{(d-z_k)^2}{v_2}
   \right]
   \left[
   \frac{z_r^2}{v_1}+\frac{(d-z_r)^2}{v_2}
   \right] \\
  & = & 
   \frac{b_N-a_N}{4d^2} \sum_{k=1}^N 
   \left[
   \frac{z_k^2}{v_1}+\frac{(d-z_k)^2}{v_2}
   \right]^2+
    \frac{a_N}{4d^2}
    \sum_{k=1}^N\sum_{r=1}^N 
   \left[
   \frac{z_k^2}{v_1}+\frac{(d-z_k)^2}{v_2}
   \right]
   \left[
   \frac{z_r^2}{v_1}+\frac{(d-z_r)^2}{v_2}
   \right] \\
  & \approx & \frac{b_N}{4d^2} \frac{1}{\Delta z}
     \int_0^d \left[ \frac{z^2}{v_1}+\frac{(d-z)^2}{v_2}
   \right]^2dz
   +
      \frac{a_N}{4d^2} \frac{1}{(\Delta z)^2}
   \left( \int_0^d
   \left[
   \frac{z^2}{v_1}+\frac{(d-z)^2}{v_2}
   \right]\,dz
    \right)^2 \\
   & = &\frac{b_N}{\Delta z}\, 
     \frac{d^3(3v_1^2+v_1v_2+v_2^2)}{60 v_1^2v_2^2}
     +\frac{a_N}{(\Delta z)^2}\,
       \frac{d^4(v_1+v_2)^2}{36 v_1^2v_2^2} \\
     & = & \frac{b_N}{\Delta z}\, 
     \frac{d^3(4v_1^2-7v_1v_2+4v_2^2)}{180 v_1^2v_2^2}
     + \frac{d^2(v_1+v_2)^2}{36 v_1^2v_2^2}
\eea
and we have for the variance
\beq \label{variance_landau}
   \Delta_{\tau}^2=\ov{\tau^2}-\ov \tau^2 = 
   \frac{b_N d}{\Delta z}\, 
     \frac{d^2(4v_1^2-7v_1v_2+4v_2^2)}{180 v_1^2v_2^2} 
\eeq
The expression for $\Delta _\tau$ is symmetric with respect to $v_1$ and $v_2$, which reflects the fact that the induced signal on the electrode at $z=0$ is always equal (and opposite in sign) to the signal at the electrode at $z=d$. To evaluate $b_N$ 
\beq
    b_N=\int_0^{\infty}\int_0^{\infty} ... \left[\int_0^{\infty}  \frac{n_1^2\,p(n_1,\Delta z)}{(n_1+n_2+...+n_N)^2}  dn_1\right]
    p(n_2, \Delta z)...p(n_N,\Delta z)\,dn_2...dn_N 
\eeq
we change variables according to $n=n_2+n_3+...+n_N$, i.e. $n_2=n-n_3-n_4-...-n_N$ and $dn_2= dn$ and see that the expression outside the brackets becomes equal to the the $N-1$ times self convoluted probability $p(n, \Delta z)$ which is simply  $p(n,d-\Delta z) \approx p(n,d)$. Using Eq. \ref{pn_approximation} for small values of $\Delta z$ the expression therefore becomes
\beq
   b_N = \int_0^\infty 
   \left[
   \int_0^\infty \frac{n_1^2\,p(n_1,\Delta z)}{(n_1+n)^2}dn_1
   \right] p(n,d)dn =
    \int_0^\infty 
   \left[
   \frac{\Delta z}{\lambda}\int_0^\infty \frac{n_1^2\,p_{clu}(n_1)}{(n_1+n)^2}dn_1
   \right] p(n,d) dn 
\eeq
so for the variance we finally have
\beq \label{final_variance}
   \Delta_{\tau}^2=\ov{\tau^2}-\ov \tau^2 = 
   w(d)^2\, 
      \left(\frac{4\,d^2}{180v_2^2}-\frac{7\,d^2}{180v_1v_2}+\frac{4\,d^2}{180 v_1^2} \right)
\eeq
\beq \label{general_xd}
   w(d)^2 =     \int_0^\infty 
   \left[
   \frac{d}{\lambda}\int_0^\infty \frac{n_1^2\,p_{clu}(n_1)}{(n_1+n)^2}dn_1
   \right] p(n,d) dn 
\eeq
This expression for $w(d)$ is completely general for any kind of cluster size distributions $p_{clu}(n)$ and resulting $p(n,d)$. 

%%%%%%%%%%%%%%%%%%%%%%%%%%%%%%
%%%%%%%%%%%%%%%%%%%%%%%%%%%%%

\section{\label{appendix_wd}}

Using the Landau theory we have $p_{clu}(n)$ from Eq. \ref{Landau_approximation} and therefore 
\beq
 \int_0^\infty \frac{n_1^2\,p_{clu}(n_1)}{(n_1+n)^2}dn_1 =
  \int_{n_0}^\infty\,\frac{n_0}{(n_1+n)^2}dn_1
 = \frac{n_0}{ n+n_0} 
\eeq
and with Eq. \ref{landau_probability} we get
\beq \label{bN}
   w(d)^2 
    =  n_0\frac{d}{\lambda}\,
   \int_0^\infty\frac{\,p(n,d)}{n+n_0} dn 
     =  
   \int_0^\infty\frac{\,L(z+\gamma-1-\ln d/\lambda)}{z+\lambda /d} dz 
\eeq
Using Eq. \ref{Landau_low} for $L(x)$ we have
\beq \label{wd_exact}
   w(d)^2
   = \int_0^\infty e^{-t\pi/2}
   \left[
   \frac{1}{2}\sin(ft) 
   - \frac{1}{\pi}\sin (ft) \mbox{SinIntegral}(t\,\lambda/d)
  - \frac{1}{\pi}\cos (ft) \mbox{CosIntegral}(t\lambda/d)
   \right] dt
\eeq
with
\beq
   f=1-\gamma+\lambda/d-\ln t+\ln d/\lambda
\eeq
The integrand is 'damped' by the exponential decay where beyond $t=10$ the integrand will be negligible. For small values of $\lambda/d$ we can use SinIntegral$(x)\approx x$ and CosIntegral$(x)\approx \gamma + \ln x$ and we get 
\beq
 w(d)^2 \approx \int_0^\infty e^{-t\pi/2}
   \left[
    \frac{1}{2}\sin(ft) - \frac{1}{\pi}(1-f)\cos (ft)
    \right] dt
\eeq
\beq
   f \approx 1-\gamma-\ln t + \ln d/\lambda 
\eeq
For $d/\lambda >40$ the approximation is accurate to better than 1\% and the dependence on $b_N$ for different sensor values of the sensor thickness is only though $\ln d/\lambda$.  For very large numbers of $d/\lambda$ the expression approaches 
\beq
    w(d)^2 = \frac{1}{\ln (d/\lambda)} \qquad d/\lambda  \rightarrow  \infty
\eeq
For $d/\lambda > 40$ this expression for $w(d)$ is within 15\% of the exact expression \ref{wd_exact}. 

%%%%%%%%%%%%%%%%%%%%%%%%%%%%%%
%%%%%%%%%%%%%%%%%%%%%%%%%%%%%

\section{\label{appendix_gain1}}

For the convolution of the Landau distribution with a Gaussian we use Eq. \ref{Landau_high} and find 
\bea
   p_G(n,d) & = &  \frac{1}{G} \int_0^\infty p(m,d)\frac{1}{\sqrt{2 \pi m} }
   \exp \left( -\frac{(n/G-m)^2}{2m} \right)dm \\
       & = &  \frac{1}{G}\int_0^\infty 
       \left[
       \frac{\lambda}{n_0 d}
       \frac{1}{\pi}
       \int_0^\infty
       \exp(-t(\frac{\lambda}{n_0d}m+\gamma-1-\ln d/\lambda)-t\ln t)
       dt
       \right]\sin (\pi t)
          \frac{\exp 
   \left( 
        - \frac{(n/G-m)^2}{2m}
   \right)}{\sqrt{2 \pi m}} dm  \no  \\
   & = & 
   \frac{1}{G}\frac{\lambda}{n_0 d \pi} 
   \int_0^\infty 
   \exp \left[
   -t(\gamma-1-\ln d/\lambda)-t\ln t+n/G\left(1-\sqrt{1+\frac{2 \lambda t}{n_0 d}}
   \right)
   \right]
   \frac{1}{
   \sqrt{1+\frac{2 \lambda t}{n_0 d}}
   }
   \sin (\pi t) dt \no \\
\eea
%
%
%

%%%%%%%%%%%%%%%%%%%%%%%%%%%%%%%%%%%%
%%%%%%%%%%%%%%%%%%%%%%%%%%%%%%%%%%%%

%%%%%%%%%%%%%%%%%%%%%%%%%%%%%%%%%%%%
%%%%%%%%%%%%%%%%%%%%%%%%%%%%%%%%%%%%

\section{ \label{appendix_weightfield}}

The expression for the weighting potential of a rectangular pad of dimension $w_x, w_y$ centred at $x=y=0$ with a parallel plate separation of $d$ is given in \cite{weighting_field} as 
\beq \label{weighting_potential}
   \phi_w(x, y, z, w_x, w_y,d) =  \frac{1}{2\pi}f(x, y, z, w_x, w_y)-\frac{1}{2\pi}\sum_{n=1}^\infty
   [f(x, y, 2nd-z,w_x,w_y)- f(x, y, 2nd+z,w_x,w_y)]   \\
\eeq
\bea
   f(x, y, u, w_x, w_y) &=& 
   \arctan \left(  \frac{x_1y_1}{u\sqrt{x_1^2+y_1^2+u^2}} \right )
   +    \arctan \left(  \frac{x_2y_2}{u\sqrt{x_2^2+y_2^2+u^2}}  \right) \\
   &-&    \arctan \left(  \frac{x_1y_2}{u\sqrt{x_1^2+y_2^2+u^2}} \right )
   -   \arctan \left(  \frac{x_2y_1}{u\sqrt{x_2^2+y_1^2+u^2}}  \right) 
\eea
\beq
       x_1= x-\frac{w_x}{2} \qquad
       x_2= x+\frac{w_x}{2} \qquad
       y_1= y-\frac{w_y}{2} \qquad
       y_2= y+\frac{w_y}{2} 
\eeq
We note that
\beq
     \phi_w (x,y,z,w_x,w_y,d) = 
     \phi_w \left(\frac{x}{d}, \frac{y}{d}, \frac{z}{d}, \frac{w_x}{d}, \frac{w_y}{d}, 1\right)
\eeq
The weighting field is given by
\beq \label{weighting_field_final}
  E_w^z(x, y, z,w_x,w_y,d) = \frac{1}{2\pi}  g(x, y, z,w_x,w_y)+\frac{1}{2\pi}\sum_{n=1}^\infty [ g(x, y, 2nd+z,w_x,w_y)+ g(x, y, 2nd-z,w_x,w_y)]  
\eeq
with
\bea
      g(x, y, u,w_x,w_y) & = & \frac{x_1y_1(x_1^2+y_1^2+2u^2)}{(x_1^2+u^2) (y_1^2+u^2)\sqrt{x_1^2+y_1^2+u^2}} 
    +\frac{x_2y_2(x_2^2+y_2^2+2u^2)}{(x_2^2+u^2)(y_2^2+u^2)\sqrt{x_2^2+y_2^2+u^2}} \no \\
  &-&  \frac{x_1y_2(x_1^2+y_2^2+2u^2)}{(x_1^2+u^2)(y_2^2+u^2)\sqrt{x_1^2+y_2^2+u^2}} 
 -\frac{x_2y_1(x_2^2+y_1^2+2u^2)}{(x_2^2+u^2)(y_1^2+u^2)\sqrt{x_2^2+y_1^2+u^2}} 
\eea
and it holds that
\beq
     E_w^z(x,y,z,w_x,w_y,d) = 
     \frac{1}{d} E^z_w \left( 
     \frac{x}{d},\frac{y}{d},\frac{z}{d},\frac{w_x}{d},\frac{w_y}{d},1
     \right)
\eeq

\clearpage

%\section{\label{tables_section}}

\begin{table}
\centering
\begin{tabular}{|c|c|c|c|c|}
\hline
 $w/d $ & $c_{22}$ &$ c_{12}$ & $c_{11}$ &      $c_{11}+ c_{12} + c_{22}   $      \\
\hline
 $ 0 $ & $0$ &$0$ & $0$ & $0$ \\
 \hline
 $0.01$ & $6.13\times 10^{-12}$ &$ -2.88\times 10^{-9}$ & $3.44\times 10^{-7}$ & $3.41\times 10^{-7}$ \\
$0.1 $& $6.05\times 10^{-8} $&$ -2.75\times 10^{-6}$ &$ 3.18\times 10^{-5} $&$ 2.91\times 10^{-5} $\\
$0.2$ & $9.28\times 10^{-7} $&$ -2.06\times 10^{-5}$ &$ 1.17\times 10^{-4} $& $9.68\times 10^{-5} $\\
$0.25$ & $2.2\times 10^{-6}$ &$ -3.88\times 10^{-5}$ &$ 1.74\times 10^{-4} $&$1.37\times 10^{-4}$ \\
$0.5 $ & $2.77\times 10^{-5}$ &$ -2.44\times 10^{-4} $&$ 5.5\times 10^{-4}$ &$3.33\times 10^{-4}$ \\
$ 1. $& $2.1\times 10^{-4}$ &$ -1.04\times 10^{-3} $& $1.33\times 10^{-3}$ &$ 4.99\times 10^{-4} $\\
 $1.5$ & $4.5\times 10^{-4}$ &$ -1.78\times 10^{-3}$ &$ 1.81\times 10^{-3} $& $4.86\times10^{-4}$ \\
$ 2. $& $6.13\times 10^{-4}$ &$ -2.18\times 10^{-3} $&$ 2.\times 10^{-3}$ &$ 4.34\times10^{-4}$ \\
$ 3.$ & $7.13\times 10^{-4}$ &$ -2.31\times 10^{-3} $&$ 1.94\times 10^{-3}$ &$ 3.41\times10^{-4}$ \\
$ 4.$ & $6.83\times 10^{-4}$ &$ -2.14\times 10^{-3}$ &$ 1.74\times 10^{-3} $& $2.77\times10^{-4}$ \\
$ 5. $& $6.26\times 10^{-4}$ &$ -1.93\times 10^{-3} $& $1.54\times 10^{-3} $& $2.32\times10^{-4}$ \\
$10 $&$4.\times 10^{-4} $& $-1.2\times 10^{-3} $&$ 9.27\times 10^{-4}$ &$1.27\times 10^{-4}$ \\
$ 20$ &$ 2.24\times 10^{-4} $& $-6.64\times 10^{-4}$ & $5.06\times 10^{-4}$ & $ 6.61\times 10^{-5}$ \\
$50 $& $9.56\times 10^{-5}$ & $-2.82\times 10^{-4} $&$ 2.13\times 10^{-4}$ & $ 2.71\times 10^{-5}$ \\
%100 & 4.89\times 10^{-5} & -1.44\times 10^{-4} & 1.09\times 10^{-4} & 1.37\times 10^{-5} \\ 
\hline
$ \infty$ & 0 & 0 & 0 & 0 \\
\hline
\end{tabular}
\caption{Coefficients $c_{11}, c_{12}, c_{22}$ from Eq. \ref{weighting_field_resolution1} for different vales of $w/d$, where $w$ is the size of the square pixel and $d$ is the thickness of the sensor.  \label{ctable}}
\end{table}
\begin{table} 
\centering
\begin{tabular}{|c|c|c|c|c|}
\hline
 $ w/d $&$ k_{22}$ & $k_{12}$ & $k_{11}$ &$ k_{11} + k_{12} + k_{22} $\\
 \hline
 0 & $0 $&$0 $& $   \frac{1}{12} = 8.33\times 10^{-2} $&$\frac{1}{12} = 8.33\times 10^{-2}$ \\
 \hline
 0.01 & $8.43\times 10^{-8} $&$ -6.43\times 10^{-5} $& $8.33\times 10^{-2} $&$8.32\times 10^{-2}$ \\
 0.1 &$ 5.37\times 10^{-5}$ &$ -2.82\times 10^{-3}$ &$ 8.05\times 10^{-2}$ & $7.77\times 10^{-2} $\\
0.2 & $3.05\times 10^{-4} $&$ -7.32\times 10^{-3} $& $7.57\times 10^{-2}$ & $ 6.87\times 10^{-2} $\\
 0.25 & $5.13\times 10^{-4}$ &$ -9.62\times 10^{-3}$ & $7.32\times 10^{-2} $ & $6.41\times 10^{-2}$ \\
0.5 & $2.17\times 10^{-3} $&$ -1.94\times 10^{-2}$ & $6.18\times 10^{-2}$ &  $4.46\times 10^{-2}$ \\
 1. & $6.39\times 10^{-3}$ & $-2.96\times 10^{-2} $& $4.73\times 10^{-2}$ &$ 2.41\times  10^{-2} $\\
 1.5 & $9.82\times 10^{-3}$ &$ -3.36\times 10^{-2}$ &$ 3.99\times 10^{-2}$ & $1.62\times  10^{-2}$ \\
 2. &$ 1.22\times 10^{-2}$ &$ -3.53\times 10^{-2}$ &$ 3.58\times 10^{-2}$ & $1.28\times  10^{-2} $\\
 3. & $1.51\times 10^{-2} $&$ -3.67\times 10^{-2} $& $3.15\times 10^{-2}$ & $9.86\times  10^{-3}$ \\
 4. &$ 1.68\times 10^{-2} $&$ -3.74\times 10^{-2} $& $2.92\times 10^{-2}$ & $8.61\times  10^{-3}$ \\
 5. &$ 1.78\times 10^{-2} $&$ -3.77\times 10^{-2} $&$ 2.78\times 10^{-2}$ & $7.92\times  10^{-3}$ \\
 10 & $2.\times 10^{-2} $&$ -3.83\times 10^{-2} $& $2.5\times 10^{-2} $& $ 6.68\times 10^{-3}$ \\
 20 & $2.12\times 10^{-2} $&$ -3.86\times 10^{-2} $&$ 2.35\times 10^{-2}$ & $ 6.19\times 10^{-3}$ \\
50 & $2.29\times 10^{-2} $& $-3.84\times 10^{-2} $&$ 2.2\times 10^{-2} $&  $6.44\times 10^{-3} $\\
% 100 & 2.63\times 10^{-2} & -3.77\times 10^{-2} & 1.96\times 10^{-2} & 8.23\times 10^{-3} \\
   \hline
  $ \infty $&$ \frac{4}{180}=2.2 \times 10^{-2}$ & $-\frac{7}{180}=-3.89 \times10^{-2} $& $\frac{4}{180}= 2.2 \times10^{-2}$ &$ \frac{1}{180}=5.56 \times 10^{-3} $\\
   \hline
\end{tabular}
\caption{Coefficients $k_{11}, k_{12}, k_{22}$ from Eq. \ref{weighting_field_resolutionk} for different vales of $w/d$, where $w$ is the size of the square pixel and $d$ is the thickness of the sensor. \label{kktable}}
\end{table}

\begin{table} 
\centering
\begin{tabular}{|c|c|c|c|c|c|c|c|c|c|c|c|c|c|c|c|c|c|c|c|}
\hline
   w/d         & 0 & 0.1   & 0.2   & 0.3 & 0.4 & 0.5  & 1 & 1.5 & 2 & 2.5 & 3 & 4 &  5 & 10 & 20 & 30 & 40 & 50 & $\infty$ \\ \hline
   $10^3\times s_{22}$ & 0 & 0.03 & 0.12 &0.27&0.54&0.76&2.6&4.0&4.8&5.2&5.2  &4.9&4.2&2.7&1.6&1.3&1.1&1.0 &0                                                                                             \\
 \hline
\end{tabular}
\caption{Coefficients $s_{22}$ from Eq. \ref{weighting_field_resolutions} for different vales of $w/d$, where $w$ is the size of the square pixel and $d$ is the thickness of the sensor. \label{sstable}}
\end{table}

\end{document}